\documentclass[a4paper,fleqn,usenatbib]{mnras}

\usepackage[T1]{fontenc}
\usepackage{ae,aecompl}

\usepackage{graphicx}	
\usepackage{amsmath}	
\usepackage{amssymb}	
\usepackage{pdflscape} 

\usepackage{rotating}
\usepackage{threeparttable}
\usepackage{caption}
\usepackage{subcaption}

\usepackage{dcolumn}
\usepackage{longtable}
\usepackage{multirow}
\usepackage{bm}
\usepackage{textpos}
\usepackage{pdflscape}	

\usepackage{geometry}
\usepackage{tikz}
\usepackage{ifthen}
\usetikzlibrary{calc}

\title[Tertiary Tides with Eccentric Orbits]{Tertiary Tides with Eccentric Orbits}
\author[Y. Gao et. al.]{
Yan Gao,$^{1}$\thanks{E-mail: ygbcyy@star.sr.bham.ac.uk}
Tjarda Boekholt$^{2}$,
Devismita Panda$^{3}$,
Tatsuya Akiba$^{4}$,
\newauthor
Silvia Toonen $^{5}$
\\
$^{1}$Institute for Gravitational Wave Astronomy \& School of Physics and Astronomy, \\
University of Birmingham, Edgbaston, Birmingham B15 2TT, UK\\
$^{2}$NASA Ames Research Center, Moffett Field, 94035, CA, USA\\
$^{3}$Laboratoire d’Astrophysique de Bordeaux, Univ. Bordeaux, CNRS,\\
B18N, All\'{e}e Geoffroy Saint-Hilaire, 33615 Pessac, France\\
$^{4}$JILA and Department of Astrophysical and Planetary Sciences, CU\\
Boulder, Boulder, CO 80309, USA\\
$^{5}$Anton Pannekoek Institute for Astronomy, University of Amsterdam, 1090 GE Amsterdam, The Netherlands\\
}

\date{Accepted XXX. Received YYY; in original form ZZZ}

\pubyear{2023}

\begin{document}
\label{firstpage}
\pagerange{\pageref{firstpage}--\pageref{lastpage}}
\maketitle

\begin{abstract}

Within hierarchical triple stellar systems, there exists a tidal process unique to them, known as tertiary tides. In this process, the tidal deformation of a tertiary in a hierarchical triple drains energy from the inner binary, causing the inner binary's orbit to shrink. Previous work has uncovered the rate at which tertiary tides drain energy from inner binaries, as a function of orbital and tidal parameters, for hierarchical triples in which the orbits are all circular and coplanar. However, not all hierarchical triples have orbits which are circular and coplanar, which requires an understanding of what happens when this condition is relaxed. In this paper, we study how eccentricities affect tertiary tides, and their influence on the subsequent dynamical evolution of the host hierarchical triple. We find that eccentricities in the outer orbit undergo tidal circularisation as quickly as binary tidal synchronisation, and are therefore trivial, but that eccentricities in the inner binary completely change the behaviour of tertiary tides, draining energy from the outer orbit as well as the inner orbit. As with the circular orbit case, tertiary tides become significant when the tertiary is large enough to come close to filling its Roche Lobe, and dominate tidal evolution when interactions between the inner binary pair are weak. Empirical equations that approximate this behaviour are provided for ease of implementing this process in other stellar evolution codes, and the implications of these results are discussed.  

\end{abstract}





\section{Introduction}

When the tertiary of a hierarchical triple is an extended body which is deformed by the tidal field of the inner binary, and when the strength of this tidal field varies periodically due to the orbital motion of the inner binary, a tidal process known as tertiary tides (TTs) comes into play, shrinking the inner binary orbit.

Indications of such a process were first observed in HD181068, a coplanar hierarchical triple with a double dwarf star inner binary and a red giant tertiary, in which the tertiary exhibited light curve behaviours which could not be explained by conventional stellar evolution models \citep{2011Sci...332..216D}. To explain these light curves, \cite{2013MNRAS.429.2425F} developed a tidal model, which also predicted that the tidal distortion of the tertiary will result in the inner binary losing orbital energy. In the case of HD181068, assuming that the tides behave as short wavelength oscillation waves in a closed system, the inner binary lost orbital energy very quickly in comparison to stellar evolutionary time-scales. Later, \cite{2018MNRAS.479.3604G} studied this process from first principles, developing an alternative model in the process, and coined the term ``tertiary tides". In their work, it was established that the rate at which orbital energy is extracted from the inner binary scales in a simple way with certain orbital and stellar parameters of the system. Further work \citep{2020MNRAS.491..264G} expanded on this idea, obtaining how these parameters influenced the orbital evolution of the host hierarchical triple in a simple functional form for triples with coplanar and circular orbits, which this paper will expand upon. Despite ongoing attempts to corroborate these findings observationally \citep{2023MNRAS.521.2114G} being temporarily inconclusive, such scaling functions have implications for triple stellar evolution, and are therefore required as input for triple stellar evolution codes \citep[e.g.][]{2017AAS...22932605T}. Previous work involving these codes \citep{2023MNRAS.518..526G} has already established that this process plays a role in the creation of Barium stars, and their impact on the formation of other celestial objects is sure to follow.

However, despite the prominent role they play, our understanding of { how quickly TTs shrink inner binary orbits} is still incomplete. { Secular evolution approximations are difficult to implement for this problem, as averaged orbits do not retain the tidal forcing mechanism behind TTs, and analytical attempts at calculating orbit-averaged energy extraction rates (such as those attempted by \citealt{2018MNRAS.479.3604G}) have their limitations. Instead, previous studies \citep{2020MNRAS.491..264G} have resorted to numerically integrating a collection of systems undergoing TTs to find trends of how quickly the inner binary orbit shrinks as a function of system parameters. They} found that, for hierarchical triples with orbits that are circular and coplanar, the rate at which the inner binary shrinks can be approximated by the empirical equation

\begin{equation}
\begin{split}
-\frac{ 1  }{  a_{\rm 1}   } \frac{ {\rm d}a_{\rm 1} } { {\rm d}t }&=\left(2.22{\times}10^{-8}{\rm yr}^{-1}\right)    \frac{4q}{\left(1+q\right)^{2}}\left(\frac{R_{\rm 3}}{100{\rm R}_{\odot}}\right)^{5.2} \\
&\left(\frac{a_{\rm 1}}{0.2{\rm AU}}\right)^{4.8}\left(\frac{a_{\rm 2}}{2{\rm AU}}\right)^{-10.2}\left(\frac{\tau}{0.534{\rm yrs}}\right)^{-1.0},
\label{Eq1} 
\end{split}
\end{equation}

\noindent where $a_{\rm 1}$ and $a_{\rm 2}$ are the semimajor axes of the inner and outer orbits, $q$ is the inner binary mass ratio, $R_{\rm 3}$ is the radius of the tertiary, and $\tau$ is the {viscous} relaxation timescale of the tertiary star for a Maxwell rheology. But not all hierarchical triples are ones in which the inner and outer orbits are circular and mutually coplanar. Even more importantly, over the course of the evolution of a hierarchical triple, there are numerous effects that can potentially drive either the inner or outer orbit into an eccentric one. This is especially true of non-coplanar hierarchical triples, where Lidov-Kozai cycles (\citealt{1962P&SS....9..719L,1962AJ.....67..591K}, see also \citealt{1910AN....183..345V}) can inject eccentricity into the system - and the interplay between the eccentricity and the mutual inclinations, when coupled with tertiary tides, can be very difficult to disentangle indeed.

In this paper, we isolate the effects of eccentricities on the effects of tertiary tides, by considering the special case where the eccentricities occur within a coplanar hierarchical triple. Thus, we can study these effects in the absence of contamination from non-coplanar influences. This should provide us with a useful stepping stone with which to prepare for subsequent investigations into tertiary tidal effects in the three-dimensional regime. In \S 2 we present our methods for investigating how eccentricities affect the host hierarchical triple, while in \S 3 we provide our results and explain them, and, finally, the implications of our work will be discussed in \S 4.

\section{Methods}

\subsection{The Model}

{ For the simulations in this paper, we adopt the same model as that used in the Stage 2 Simulations of \cite{2018MNRAS.479.3604G}. The integrator we use is also identical, 8th-order Runge Kutta. The relevant equations being solved have been added to Appendix \ref{appA}, for the reader's convenience.}

\subsection{The Setup}

Consider a hierarchical triple, the inner binary of which consists of two point masses, the masses of which are $m_{\rm 1}$ and $m_{\rm 2}$, orbiting each other at a semimajor axis of $a_{\rm 1}$, with an eccentricity of $e_{\rm 1}$. The tertiary, the mass of which is $m_{\rm 3}$, orbits the centre of mass of this binary at a semimajor axis of $a_{\rm 2}$, with an eccentricity of $e_{\rm 2}$, and is initially tidally synchronised with its mean motion around the centre of mass of the inner binary to eliminate contamination from binary tidal effects to the greatest extent possible. For this work, we only consider coplanar triples, for which the inclination angle between the inner and outer orbits is $i=0$. 

We start at a set of initial conditions that will form the anchor of our simulation runs, which we refer to as the Hypothetical Scenario. In this set of initial conditions, $a_{\rm 1}=0.2$AU, $a_{\rm 2}=2$AU, and the masses are set to $m_{\rm 1}=m_{\rm 2}=0.8$M$_{\odot}$, and $m_{\rm 3}=1.6$M$_{\odot}$. The tertiary is assumed to be a red giant with a convective envelope, with a radius of $R_3=100R_{\odot}$. The viscoelastic relaxation time $\tau$ \citep[e.g.][]{2014A&A...571A..50C} is set to 0.534 years, { which was the value found for the $e_{\rm 1}=0$ case in \cite{2018MNRAS.479.3604G}. In their work, they simulated TTs using both the viscoelastic model and an equilibrium tide model corrected for fast tides, and adjusted $\tau$ by hand until the results of the two models matched.} This choice for the value of ${\tau}$ is further justified in the discussion section of this paper. The orbits are chosen to be prograde with respect to each other, for ease of logical deduction when interpreting the results, as will be demonstrated. This choice of initial conditions is selected for consistency with \cite{2020MNRAS.491..264G}.

We then proceed to individually change either $e_{\rm 1}$ or $e_{\rm 2}$ while keeping all the other parameters the same, gradually increasing both up to 0.9, and for each modified value, we simulate the tidal evolution of the hierarchical triple for $10^5$ years, during which we seek to measure how the system changes as a result of tertiary tides.

The initial conditions we use are summarised in Table \ref{sim_params}.

\begin{table}
	\centering
	\caption{Initial parameters for our simulations, both for the hypothetical scenario which we use as a starting point, as well as the range of values that each parameter was varied over.}
	\label{sim_params}
	\begin{tabular}{ccc}
		\hline
		Parameter & Hypothetical Scenario & Range\\
		\hline
                $a_1$/AU & 0.2 & - \\
                $a_2$/AU & 2.0 & - \\
                $e_1$ & 0 & 0 - 0.9 \\
                $e_2$ & 0 & 0 - 0.9 \\
                $i$ & 0 & - \\
                $m_1$/$M_{\odot}$ & 0.8 & - \\
                $m_2$/$M_{\odot}$ & 0.8 & - \\
                $m_3$/$M_{\odot}$ & 1.6 & - \\
                $R_3$/$R_{\odot}$ & 100 & - \\
                $\tau$/years & 0.534 & - \\
		\hline
	\end{tabular}
\end{table}

\subsection{Error Estimation \& Stepsize control}

Since the calculations in this paper are done for eccentric orbits, which require a smaller integration stepsize at periastron, we scale our stepsize to be proportional to $\frac{r^{1.5}}{\sqrt{1+e_{\rm 1}}}$, where $r$ is the distance between the inner binary pair at any given moment, and $n$ is a real exponent. This is equivalent to a conventional ${\rm d}t=\sqrt{{\rm d}r^3/{\mu}}$ prescription, with an additional mechanism that further decreases the stepsize for greater inner binary eccentricities.

For the purposes of error estimation, we run our code while setting $R_3$=0. This is physically equivalent to treating all 3 bodies as point masses, but the results should include all the errors introduced by our tidal algorithm. Since the dynamical evolution of 3 point masses should conserve total energy, any deviation from perfect total energy conservation should be purely the result of our code. We find that, despite the enhanced temporal resolution for greater inner binary eccentricities, the errors increase monotonically with said eccentricity, increasing up to $6{\times}10^{34}$J over $10^{5}$ years. By comparison, the errors found by \cite{2020MNRAS.491..264G} for a similar run using our earlier code for circular orbits leads to an energy discrepancy of $2{\times}10^{33}$J. { To put this in context, the inner binary has an initial orbital energy of $-2.8{\times}10^{39}$J for our simulations in Table \ref{sim_params}, and therefore our relative energy errors are on the order of $10^{-5}$.}

\section{Results}

The results of our simulations mainly concern two scenarios, one in which the inner binary is initially circular, and the outer orbit is eccentric, and the other in which the inner binary is initially eccentric, and the outer orbit is circular.


\begin{figure*}
     \centering
    \begin{subfigure}[t]{0.49\textwidth}
        \raisebox{-\height}{\includegraphics[scale=0.29, angle=0, trim= 2.2cm 0cm 0.1cm 0cm]{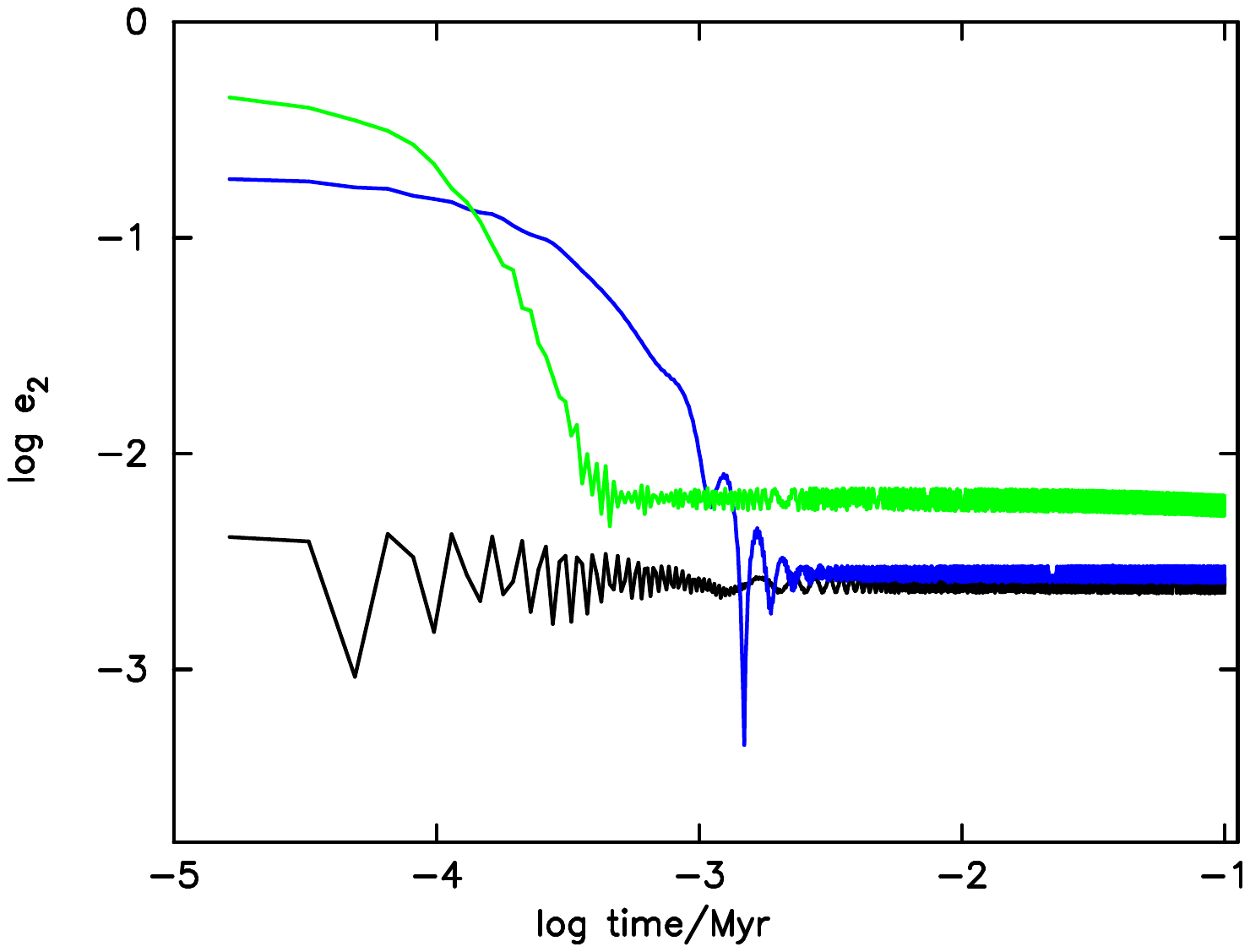}}
    \end{subfigure}
    \hfill
    \begin{subfigure}[t]{0.49\textwidth}
        \raisebox{-\height}{\includegraphics[scale=0.29, angle=0, trim= 2.2cm 0cm 0.1cm 0cm]{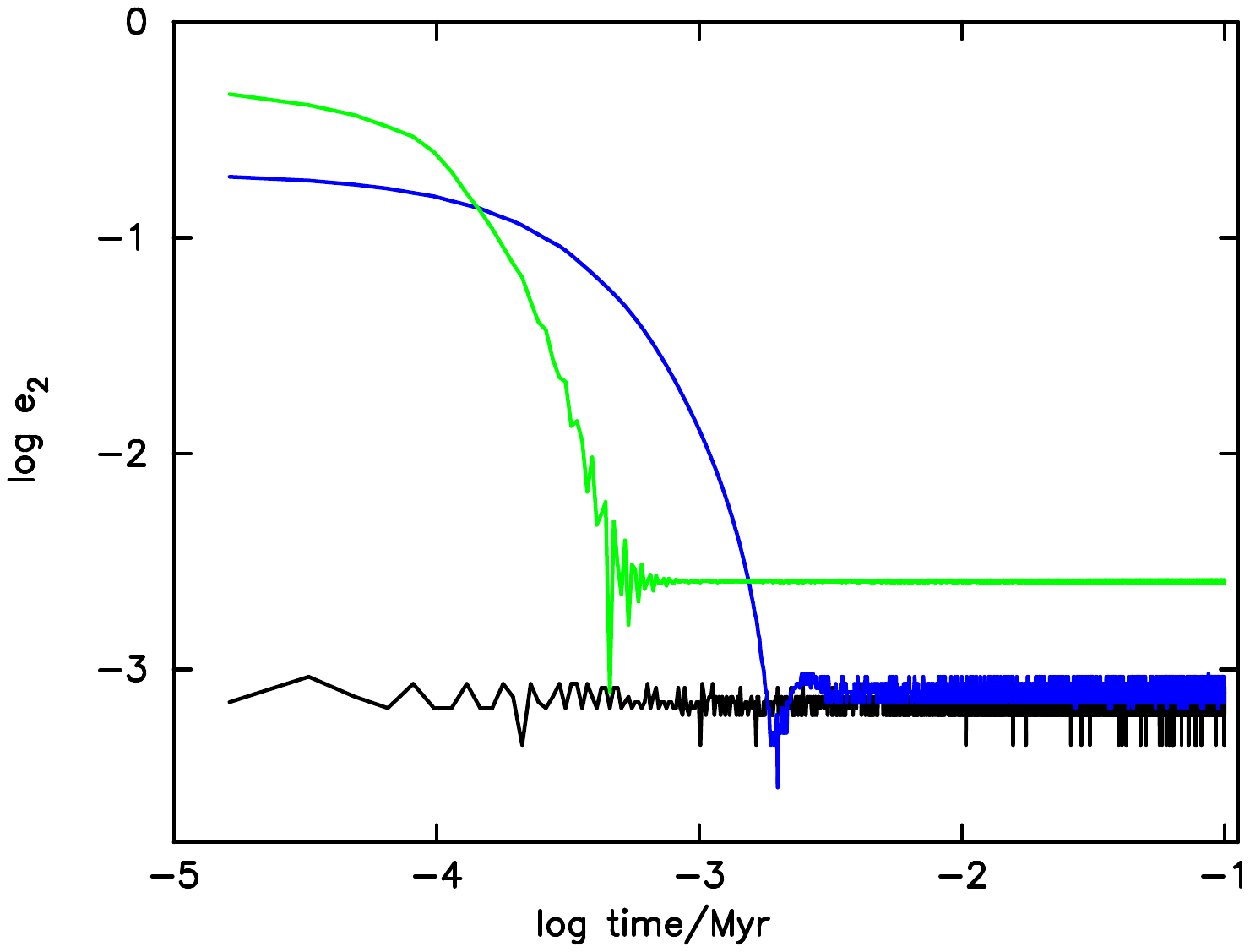}}
    \end{subfigure}
     \centering
    \begin{subfigure}[t]{0.49\textwidth}
        \raisebox{-\height}{\includegraphics[scale=0.29, angle=0, trim= 2.2cm 0cm 0.1cm 0cm]{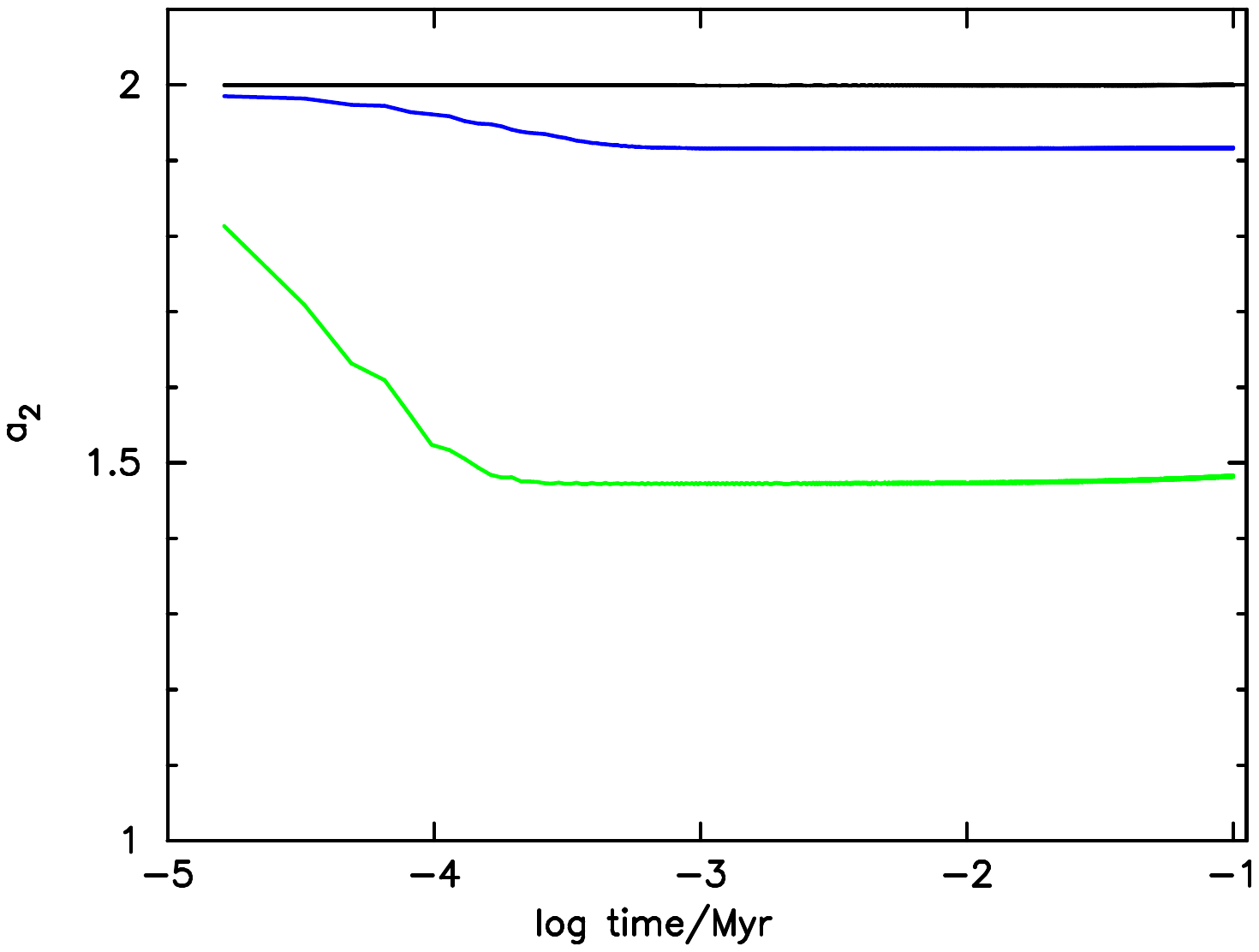}}
    \end{subfigure}
    \hfill
    \begin{subfigure}[t]{0.49\textwidth}
        \raisebox{-\height}{\includegraphics[scale=0.29, angle=0, trim= 2.2cm 0cm 0.1cm 0cm]{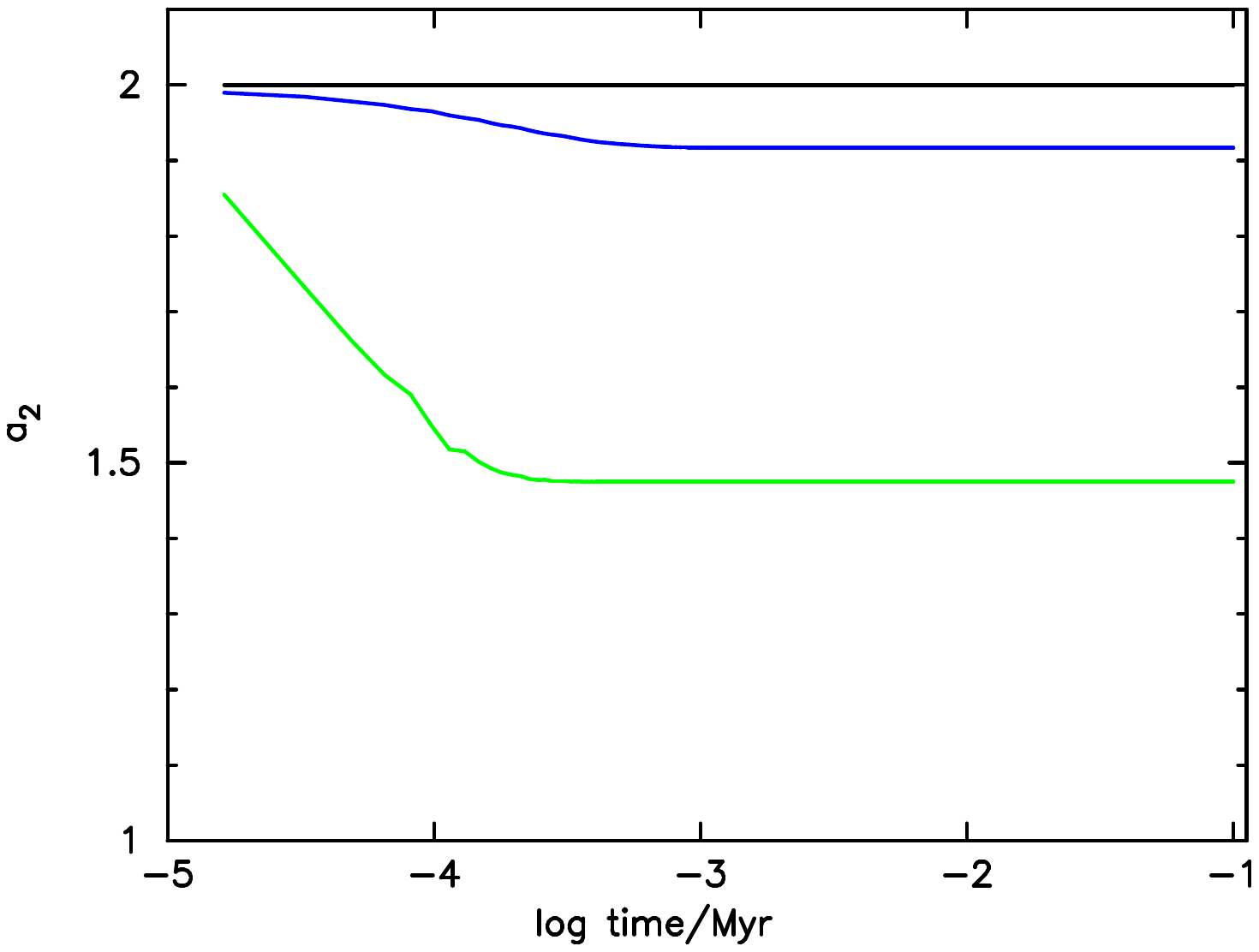}}
    \end{subfigure}
    \caption{Evolution of $e_{\rm 2}$ (first row) and $a_{\rm 2}$ (second row) for simulations in which the initial eccentricity of the outer orbit is varied. The left column shows the evolution of our Hypothetical Scenario, while the right column corresponds to an equivalent binary which is obtained by replacing the inner binary in the Hypothetical Scenario with a point mass of $m_{\rm 1}+m_{\rm 2}$. The black, blue and green lines represent the evolution of initial values of $e_{\rm 2}$ of 0.0, 0.2, and 0.5, respectively. The small non-zero residual eccentricity at the end of the circularisation process is a known feature of tertiary tides for circular orbits. The only other major difference between the left and right columns appears to be some small extra oscillation of the curves in the triple case, as would be expected from the perturbation by the inner binary. It is obvious that tidal cirularisation takes over very quickly, and that the process is dominated by conventional binary tidal synchronisation.}
    \label{fig1}
\end{figure*}

\subsection{Eccentric Outer Orbits}

We find that, for any outer orbit that is initially eccentric, the eccentricity vanishes very quickly, in a matter of a few hundred years for $\tau=$0.534 years, as shown in Figure \ref{fig1}. During this process, the angular momentum of the outer orbit is conserved, and therefore the final semimajor axis of the outer orbit always shrinks to $a_{\rm 2}\left(1-e_{\rm 2}^{\rm 2}\right)$, with some minor deviation due to the energy and angular momentum input from the tertiary's rotation.

This circularisation process is easily explained by two-body tidal circularisation. When the outer orbit is eccentric, the inner binary can be seen as a single tidal companion to the extended tertiary, and due to the large radius of the giant tertiary, its orbit undergoes tidal circularisation quickly. To demonstrate that this is indeed the case, we replace the inner binary with a single star of mass $m_{\rm 1}+m_{\rm 2}$, and repeat the simulation. The results are also shown in Figure \ref{fig1}, where it can be seen that the effects are practically identical to those induced by the inner binary.


\subsection{Eccentric Inner Orbits}

\begin{figure*}
     \centering
    \begin{subfigure}[t]{0.48\textwidth}
        \raisebox{-\height}{\includegraphics[scale=0.29, angle=0, trim= 2.2cm 0cm 0.1cm 0cm]{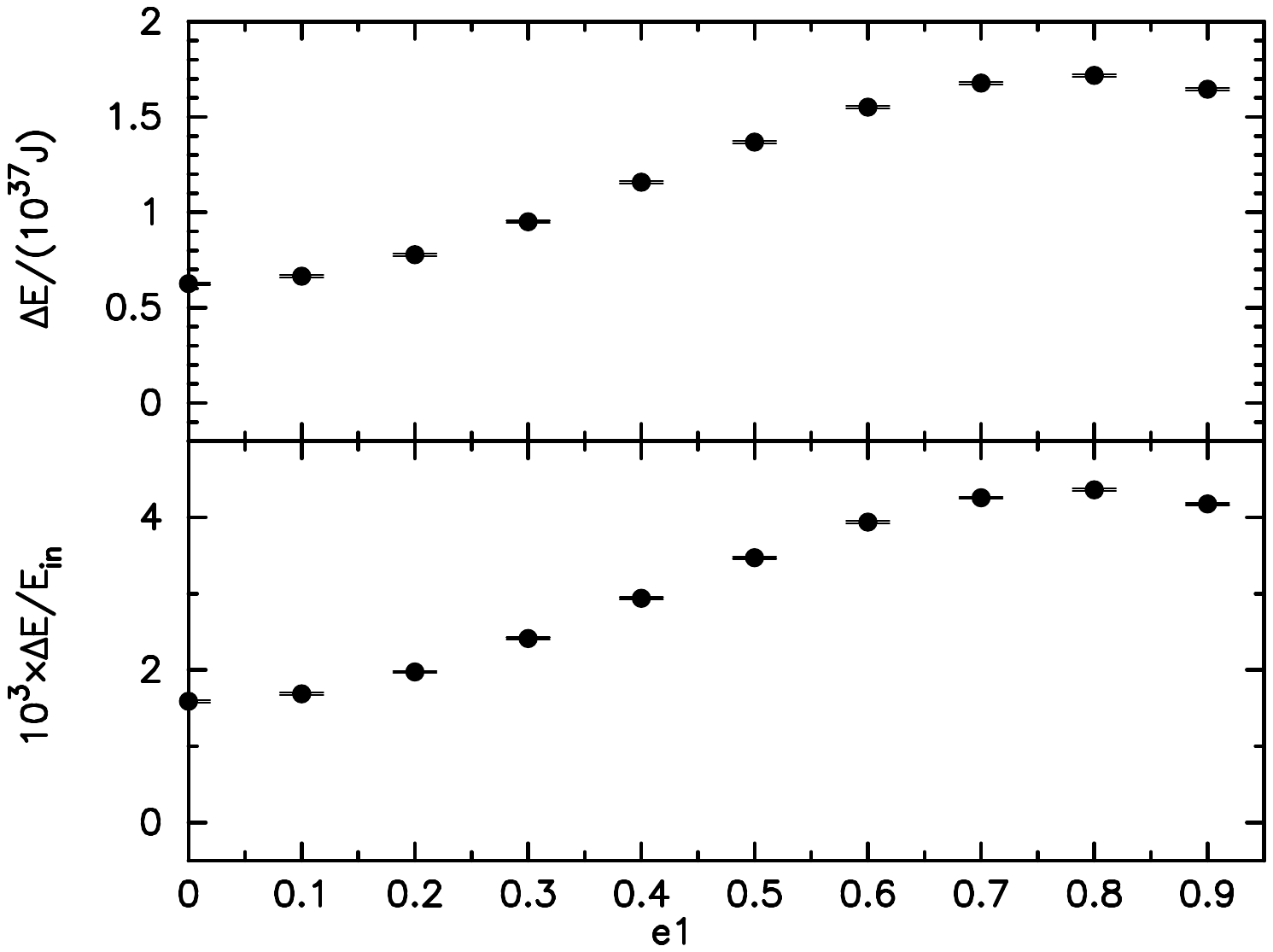}}
    \end{subfigure}
    \hfill
    \begin{subfigure}[t]{0.48\textwidth}
        \raisebox{-\height}{\includegraphics[scale=0.33, angle=0, trim= 2.2cm 2.8cm 0.1cm 2.3cm]{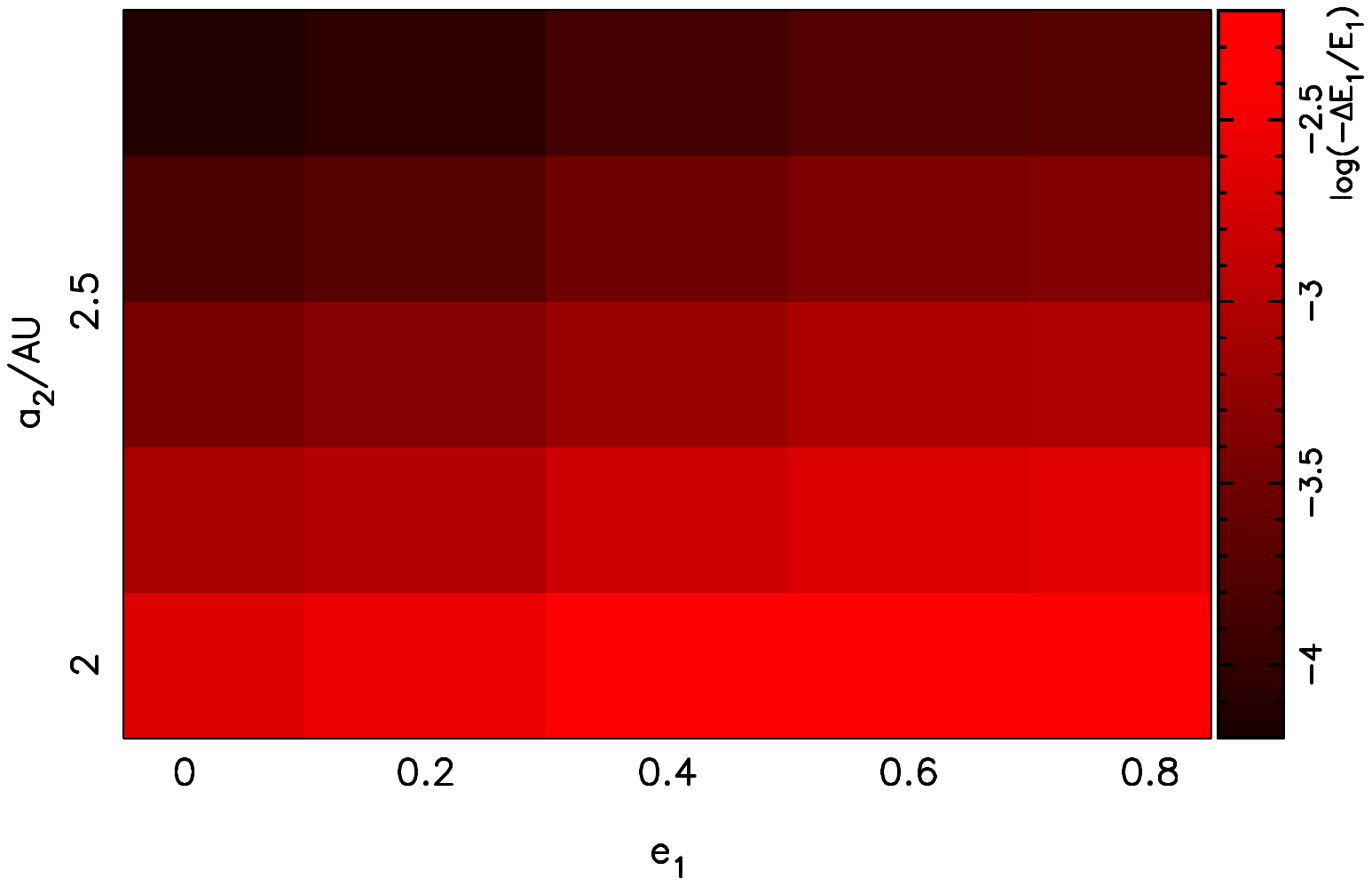}}
    \end{subfigure}
    \caption{The upper half of the left panel plots energy extracted from the Hypothetical Scenario as a function of initial inner binary orbital eccentricity, while the lower half dsiplays the same energy extraction in units of initial inner binary orbital energy. The { vertical} error bars, which are very small, indicate the systematic errors due to the algorithm ($6{\times}10^{34}$J). Since they are small, further plots will dispense with these error bars. The right panel displays a heat map of how inner binary orbital shrinkage varies over the course of $10^5$ years as a function of initial $e_{\rm 1}$ and $a_{\rm 2}$, when all other parameters conform to our Hypothetical Scenario.}
    \label{fig2}
\end{figure*}

Evolving our system for $10^5$ years under initial inner binary eccentricities ranging from 0 to 0.9, with increments of 0.1, the amount of energy extracted from the inner binary is found to be that plotted in {the left panel of} Figure \ref{fig2}. This energy extraction is found by measuring the difference in energy at the local minima as the energy value oscillates, using the same methods as \cite{2020MNRAS.491..264G}. The error bars in this figure represent an error of $6{\times}10^{34}$J both ways. Since the errors are too small to be noticeable in this plot, we dispense with these error bars in the remaining plots in this paper. Repeating this process for a range of $e_{\rm 1}$ and $a_{\rm 2}$ initial values yields the heat map of inner binary orbital shrinkage in the right panel of Figure \ref{fig2}.


\begin{figure*}
     \centering
    \begin{subfigure}[t]{0.49\textwidth}
        \raisebox{-\height}{\includegraphics[scale=0.29, angle=0, trim= 2.2cm 0cm 0.1cm 0cm]{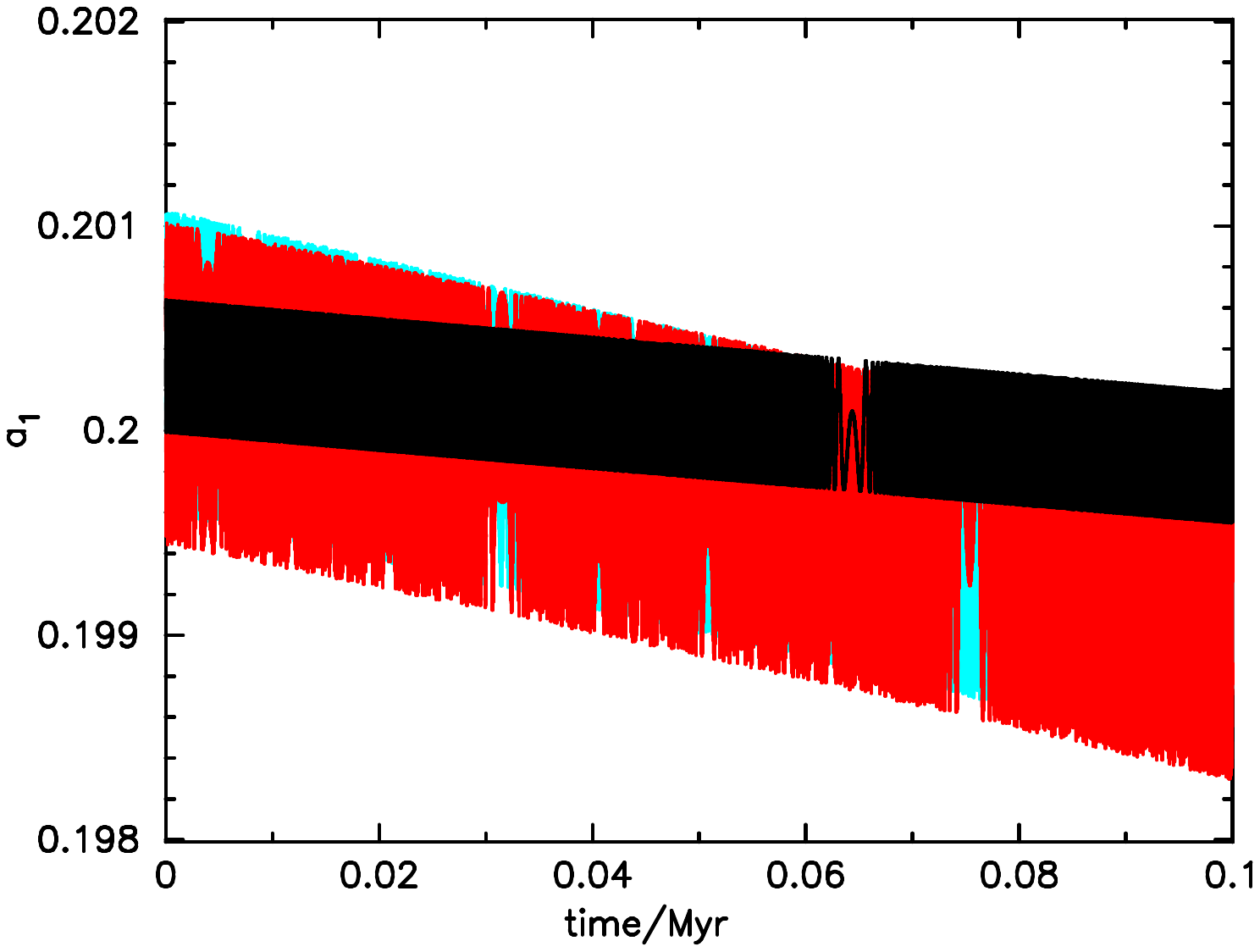}}
    \end{subfigure}
    \hfill
    \begin{subfigure}[t]{0.49\textwidth}
        \raisebox{-\height}{\includegraphics[scale=0.29, angle=0, trim= 2.2cm 0cm 0.1cm 0cm]{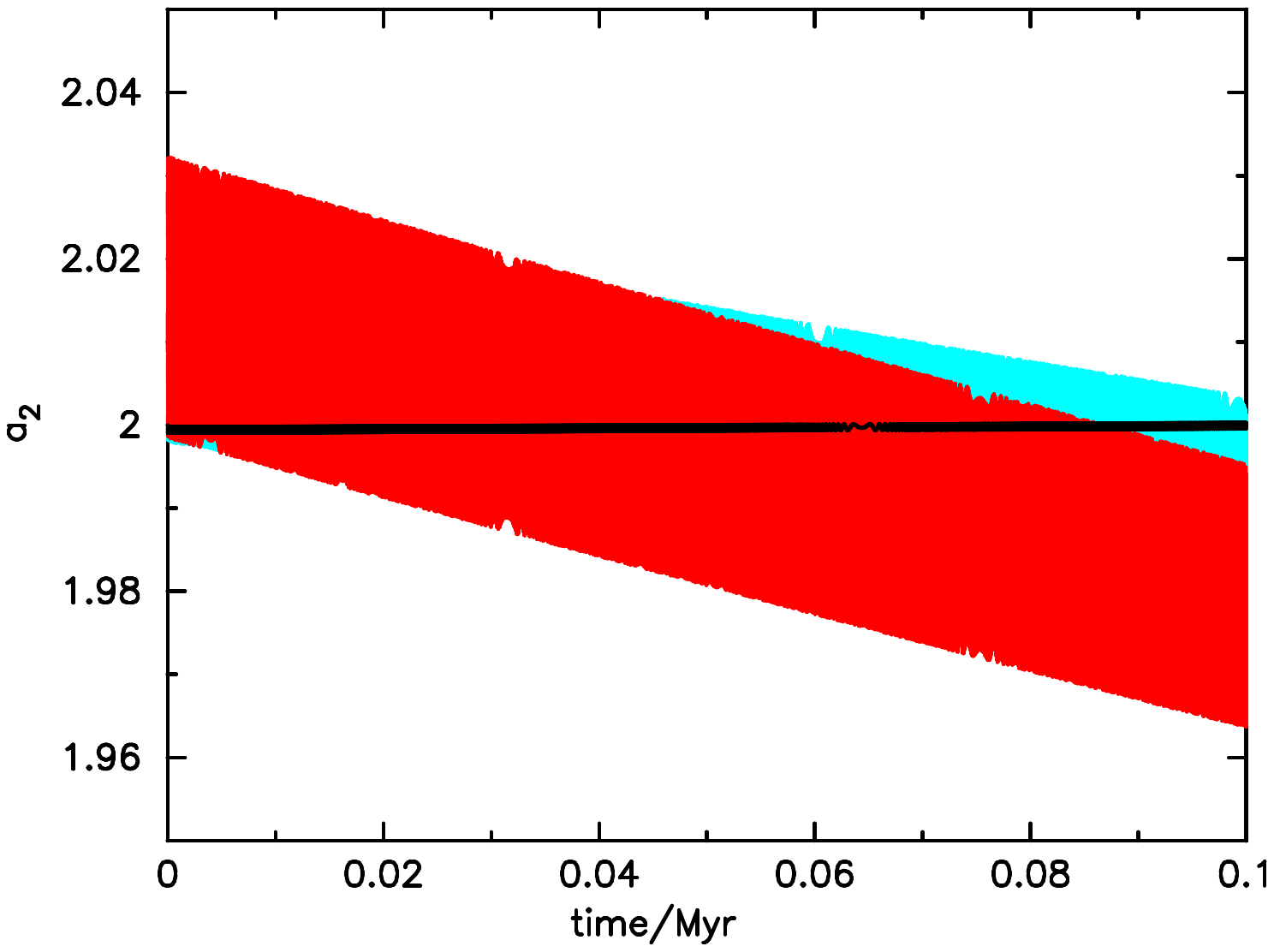}}
    \end{subfigure}
     \centering
    \begin{subfigure}[t]{0.49\textwidth}
        \raisebox{-\height}{\includegraphics[scale=0.29, angle=0, trim= 2.2cm 0cm 0.1cm 0cm]{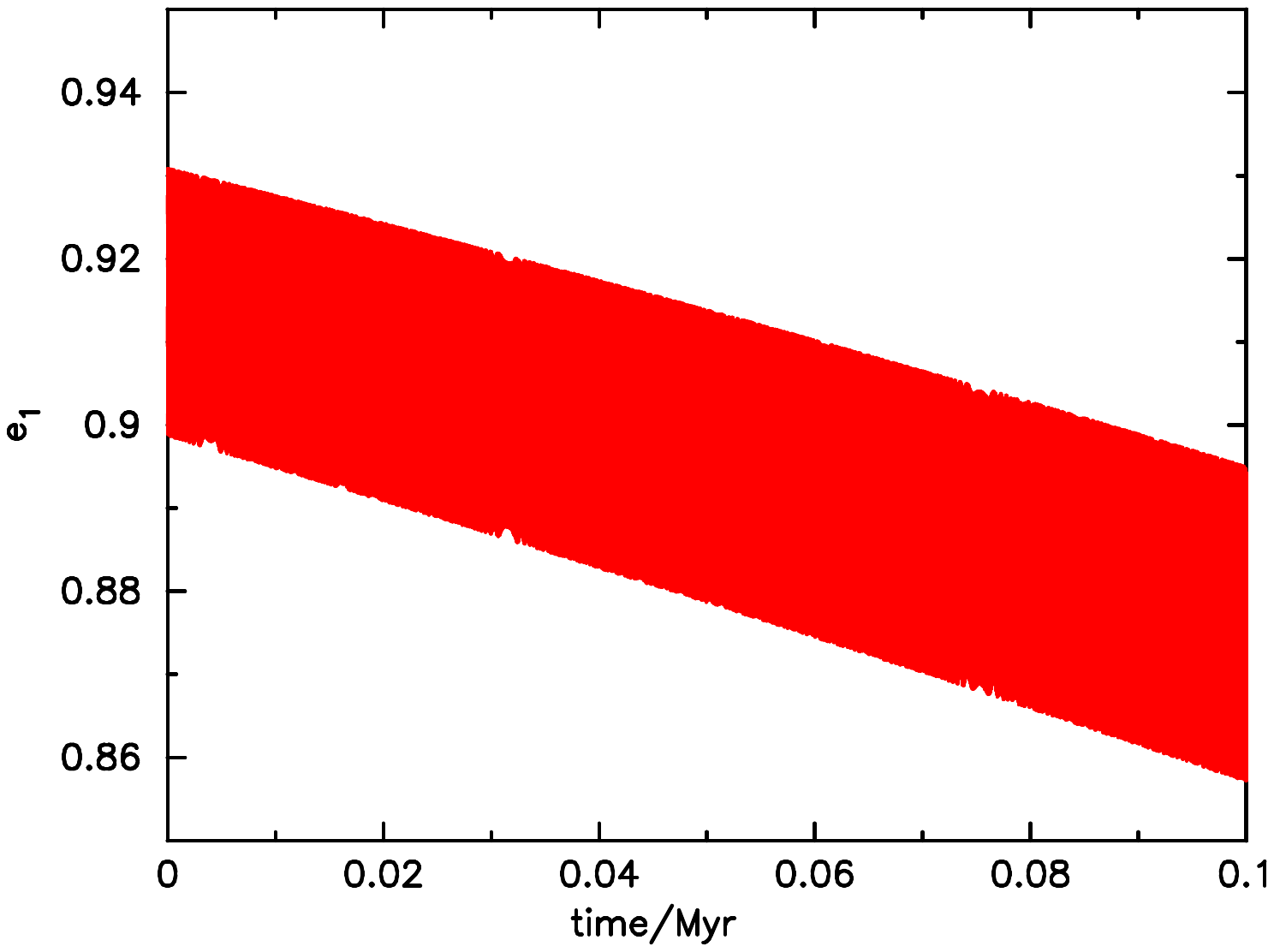}}
    \end{subfigure}
    \hfill
    \begin{subfigure}[t]{0.49\textwidth}
        \raisebox{-\height}{\includegraphics[scale=0.29, angle=0, trim= 2.2cm 0cm 0.1cm 0cm]{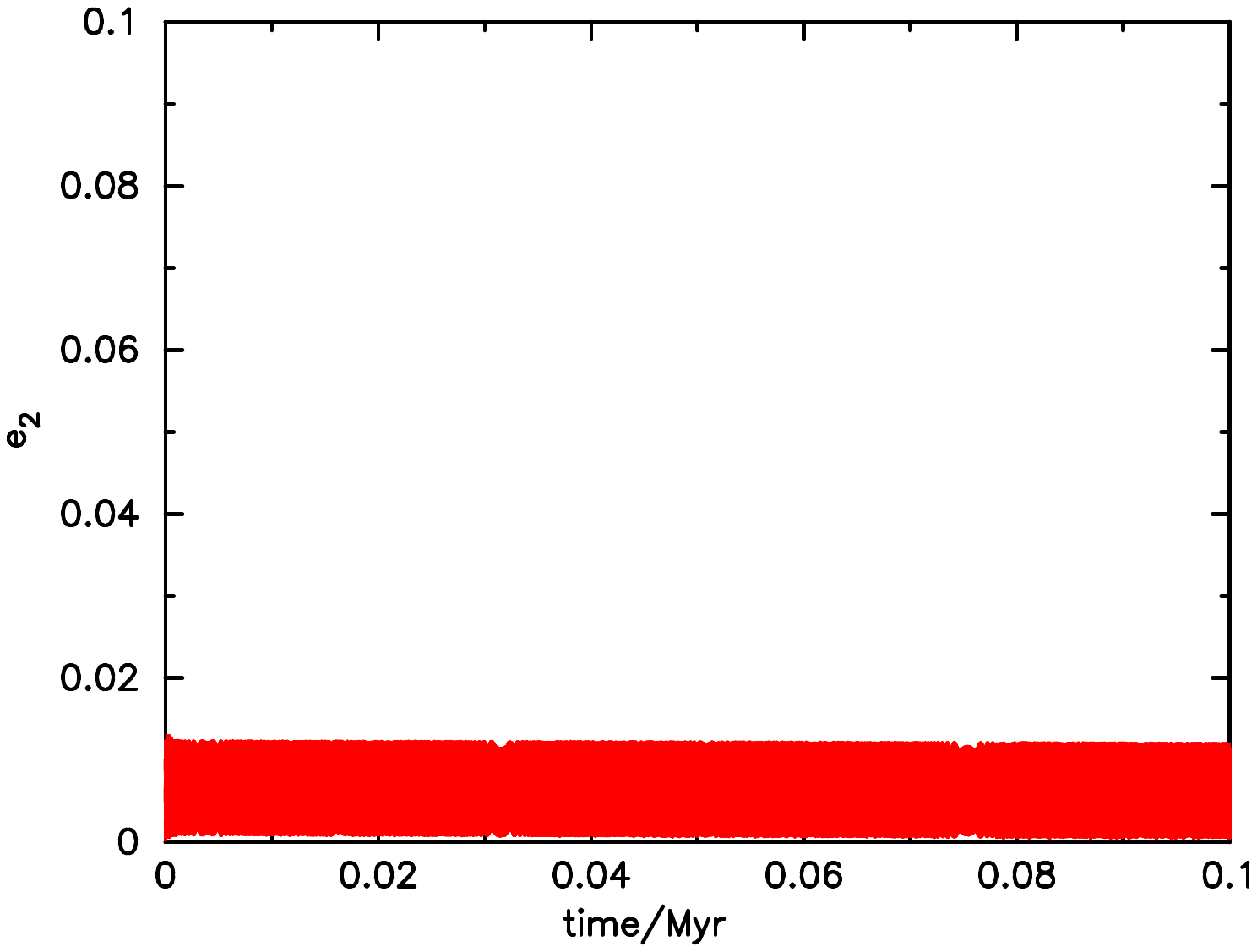}}
    \end{subfigure}
    \caption{Upper panels: evolution of the orbital semimajor axes $a_{\rm 1}$ (left) and $a_{\rm 2}$ (right) over 0.1 Myr for the Hypothetical Scenario. The black, cyan and red lines correspond to an initial $e_{\rm 1}$ value of 0, 0.8, and 0.9, respectively. Lower panels: evolution of the orbital eccentricities $e_{\rm 1}$ (left) and $e_{\rm 2}$ (right) over the same period, for the run in which the initial vallue of $e_{\rm 1}$ is 0.9. It is worth noting that both the inner and outer semimajor axes shrink, while the inner orbit's eccentricity is decreasing too, on similar timescales. Some outer orbit eccentricity is observed, which is consistent with residual eccentricity commonly seen in tertiary tidal systems.}
    \label{fig3}
\end{figure*}


As is obvious from the figure, the relation between the inner binary orbital energy extraction and initial eccentricity is not monotonic. The complexity of the issue is further compounded by the evolution of the inner binary eccentricity and outer orbit semimajor axis, which unexpectedly both decrease as a function of time, on time-scales apparently similar to that of the inner binary orbital shrinkage, as shown in Figure \ref{fig3}.

Due to the linear dependence of $\frac{1}{a_{\rm 1}}\frac{{\rm d}a_{\rm 1}}{{\rm d}t}$ on $a_{\rm 1}^{4.8}$ seen in Equation \ref{Eq1}, one would think that, for an eccentric inner orbit, the rate of energy extraction from the inner binary must behave as $\frac{1}{a_{\rm 1}}\frac{{\rm d}a_{\rm 1}}{{\rm d}t}{\propto}{\langle}r^{n}{\rangle}$, where $r$ is the distance between the inner binary pair at any given moment, and ${\langle}r^{n}{\rangle}$ is the time-averaged value of ${\langle}r^{n}{\rangle}$ over an inner orbital period, and could be calculated according to the methods of \cite{2017ApJ...837L...1M}, repeated below.

For any given moment in time within an orbit corresponding to an eccentric anomaly $E$, we can express $r$ as $r=a_{\rm 1}\left(1-e_{\rm 1}{\cos}E\right)$, and the mean anomaly $M$ as $M=E-e_{\rm 1}{\sin}E$. Thus, the integral

\begin{equation}
{\langle}r^{n}{\rangle}=\frac{1}{T}{\int}_{0}^{T}r^{n}{\rm d}t
\end{equation}

\noindent becomes

\begin{equation}
\begin{split}
{\langle}r^{n}{\rangle}&=\frac{1}{2{\pi}}{\int}_{0}^{2{\pi}}a_{\rm 1}^{\rm n}\left(1-e_{\rm 1}{\cos}E\right)^{n}{\rm d}M \\
&=\frac{1}{2{\pi}}{\int}_{0}^{2{\pi}}a_{\rm 1}^{\rm n}\left(1-e_{\rm 1}{\cos}E\right)^{n}{\rm d}\left(E-e_{\rm 1}{\sin}E\right),
\end{split}
\end{equation}

\noindent which simplifies to

\begin{equation}
{\langle}r^{n}{\rangle}=\frac{a_{\rm 1}^{\rm n}}{2{\pi}}{\int}_{0}^{2{\pi}}\left(1-e_{\rm 1}{\cos}x\right)^{n+1}{\rm d}x.
\label{rn}
\end{equation}

\begin{figure}
\includegraphics[scale=0.29, angle=0, trim= 2.2cm 0cm 0.1cm 0cm]{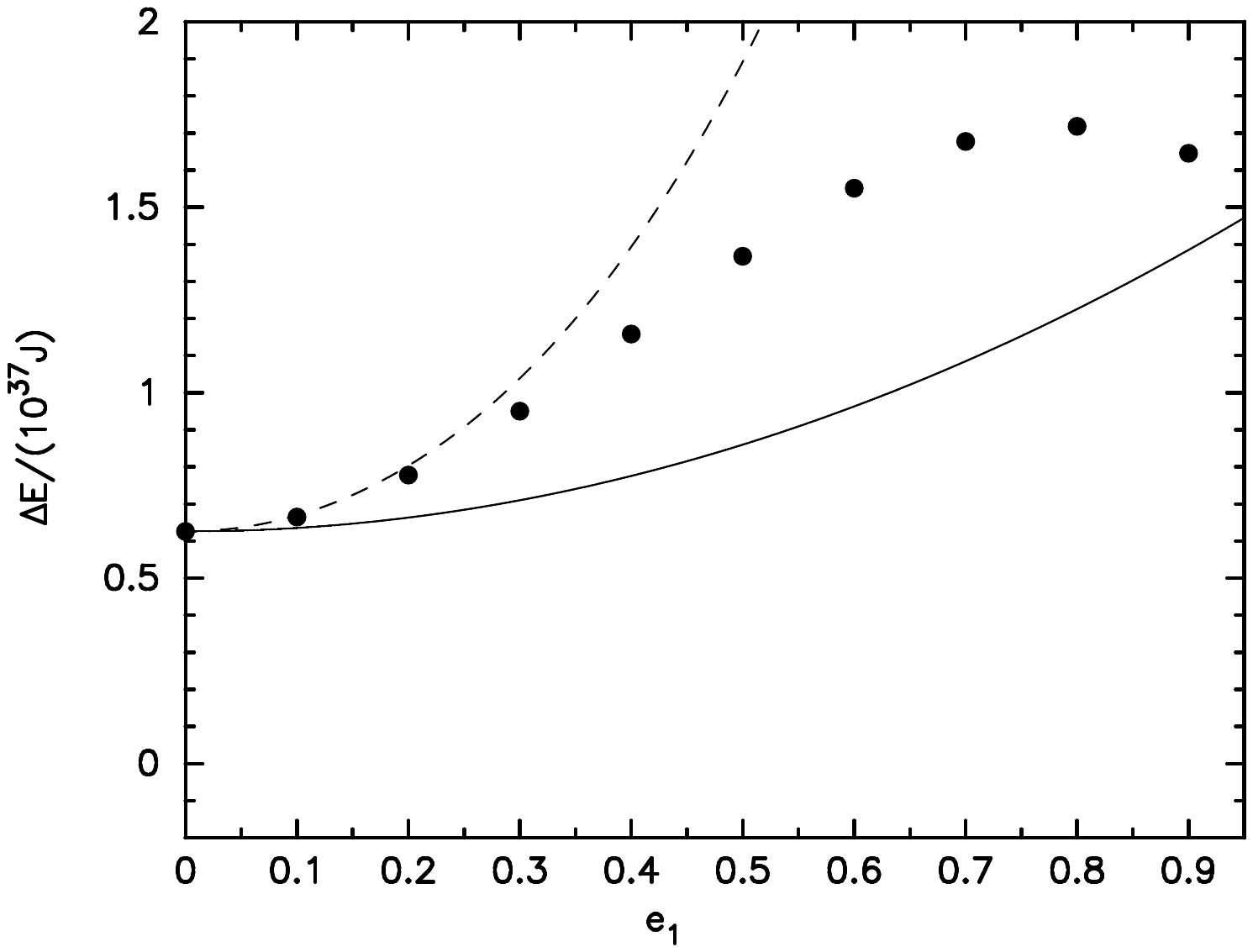}
\caption{$n=4.8$ (dashed line) and $n=2$ (solid line) curves, calculated via Equation \ref{rn}, overlaid on the upper half of the left panel of Figure \ref{fig2}. \label{fig4}}
\end{figure}

In Figure \ref{fig4}, we plot the expected curves for $n=4.8$ (motivated by the 4.8th power in our previous empirical equation) and $n=2$ (motivated by gravitational quadrupole arguments), neither of which fit the results well. Similarly, other values for $n$ also fail due to the issue that such a fit always results in a monotonic function, and thus cannot replicate the non-monotonic behaviour seen. The problems of such fits are further exacerbated when it is noted that it fails to account for the orbital shrinkage of the outer orbit, as well as the gradual decrease of $e_{\rm 1}$.

After much experimentation, we find that the best explanation of the trends we see in Figures \ref{fig2} and \ref{fig3} is that, for an eccentric inner binary orbit, the amount of energy dissipated from the inner binary drops to a fraction $1-e_{\rm 1}^{\rm 2}$ of the circular value, therefore for the inner orbit,

\begin{equation}
\begin{split}
\frac{1}{E_{\rm 1}}\frac{ {\rm d}E_{\rm 1} } { {\rm d}t }&=C\frac{{\langle}r^{4.8}{\rangle}}{a_{\rm 1}^{4.8}}\left(1-e_{\rm 1}^{\rm 2}\right) \\
&\left(\frac{a_{\rm 1}}{0.2{\rm AU}}\right)^{4.8}\left(\frac{a_{\rm 2}}{2{\rm AU}}\right)^{-10.2}\frac{4q}{\left(1+q\right)^{2}} \\
&\left(\frac{R_{\rm 3}}{100{\rm R}_{\odot}}\right)^{5.2}\left(\frac{\tau}{0.534{\rm yrs}}\right)^{-1.0},
\label{dE1dti}
\end{split}
\end{equation}

\noindent where $E_{\rm 1}$ is the orbital energy of the inner orbit, and $C=2.22{\times}10^{\rm -8}$/yr, according to our previous papers, and ${\langle}r^{4.8}{\rangle}$ is calculated, as we previously discussed, according to Equation \ref{rn}.

\begin{figure}
\includegraphics[scale=0.29, angle=0, trim= 2.2cm 0cm 0.1cm 0cm]{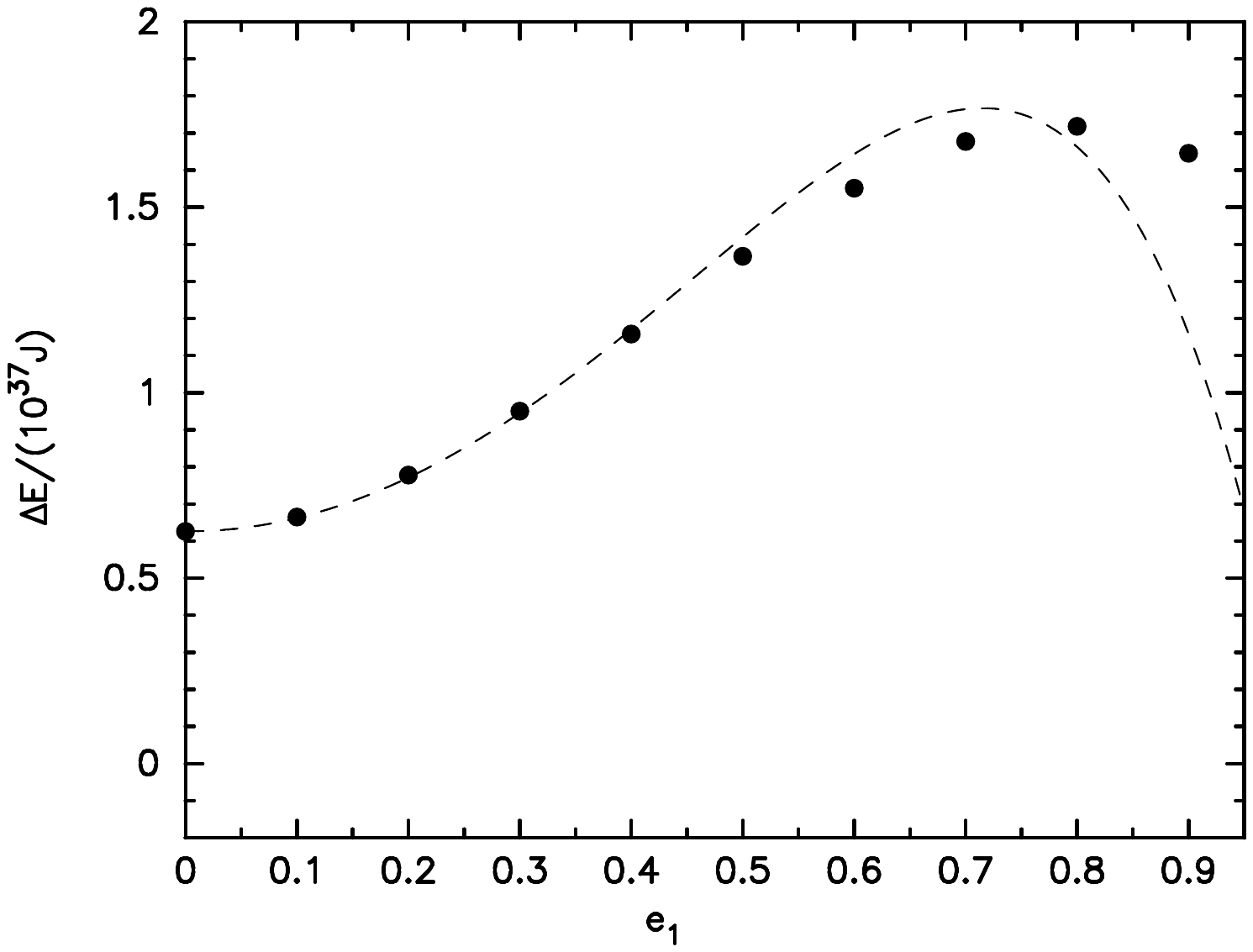}
\caption{A fraction $1-e_{\rm 1}^{\rm 2}$ of the $n=4.8$ curve overlaid on the upper half of the left panel of Figure \ref{fig2}. \label{fig5}}
\end{figure}

If we were to assume that the initial eccentricity and outer orbital semimajor axis do not change throughout the evolution, then this leads to the dashed curve shown in Figure \ref{fig5}. It can be seen that this curve matches the low initial $e_{\rm 1}$ regime quite well, while there is still some discrepancy in the high initial $e_{\rm 1}$ regime.

However, as previously shown in Figures \ref{fig2} and \ref{fig3}, the eccentricity and outer orbital semimajor axis do evolve, and to account for this evolution, we apply the following reasoning and calculations.

To begin with, why does the outer orbit shrink? The mass distribution of the inner binary averaged over its orbit, when rotating in its orbital plane around the axis passing through its centre of mass, is not rotationally symmetric. This creates a gravitational quadrupole, which the tertiary feels if the angular velocity of the outer orbit does not exactly match that of the apsidal precession of the inner binary. The strength of this interaction scales proportionally to $e_{\rm 1}^{2}$ (see derivation in Appendix \ref{appB}), which explains why the same outer orbital shrinkage is not seen in TTs with circular inner binary orbits.

If this explanation is true, then the reason the outer orbit shrinks finds a ready analogue in why the inner binary shrinks in TTs with a circular inner binary orbit. If we were to connect the inner binary pair with an ideal rigid massless rod, and prevent the inner binary from shrinking, then the inner binary would have a gravitaional quadrupole similar to that of an averaged eccentric orbit. Since the inner binary cannot shrink in this case, the energy extracted can only come from the outer orbit, but the rate at which this energy is lost must scale identically to ($q$, $a_{\rm 1}$, $a_{\rm 2}$, $R_{\rm 3}$, ${\tau}$) as the circular inner binary orbit case. Therefore, we write


\begin{equation}
\begin{split}
\frac{1}{E_{\rm 1}}\frac{ {\rm d}E_{\rm 2} } { {\rm d}t }&=C\frac{{\langle}r^{4.8}{\rangle}}{a_{\rm 1}^{4.8}}\left(Ne_{\rm 1}^{\rm 2}\right) \\
&\left(\frac{a_{\rm 1}}{0.2{\rm AU}}\right)^{4.8}\left(\frac{a_{\rm 2}}{2{\rm AU}}\right)^{-10.2}\frac{4q}{\left(1+q\right)^{2}} \\
&\left(\frac{R_{\rm 3}}{100{\rm R}_{\odot}}\right)^{5.2}\left(\frac{\tau}{0.534{\rm yrs}}\right)^{-1.0},
\label{dE2dti}
\end{split}
\end{equation}

\noindent where $N$ is of order unity, the $E_{\rm 1}$ in the first fraction is not a typo, and $E_{\rm 1}$ and $E_{\rm 2}$ are connected to $a_{\rm 1}$ and $a_{\rm 2}$ by

\begin{equation}
E_{\rm 1}=-\frac{Gm_{\rm 1}m_{\rm 2}}{2a_{\rm 1}},
\label{E1}
\end{equation}

\begin{equation}
E_{\rm 2}=-\frac{G\left(m_{\rm 1}+m_{\rm 2}\right)m_{\rm 3}}{2a_{\rm 2}}.
\label{E2}
\end{equation}

To explain the evolution of $e_{\rm 1}$, we note that angular momentum is conserved in tidal processes. Since the system that we simulate is one in which the inner and outer orbits are mutually prograde, the only way to conserve angular momentum if both orbits were to shrink is to decrease $e_{\rm 1}$. While it may be argued that the spin angular momentum of the tertiary may also play a role in conserving angular momentum, we note that the realistic amount of angular momentum that can be stored in the tertiary is insufficient to fully compensate for the decrease in total angular momentum, if $e_{\rm 1}$ were to stay constant.

\begin{figure*}
     \centering
    \begin{subfigure}[t]{0.48\textwidth}
        \raisebox{-\height}{\includegraphics[scale=0.29, angle=0, trim= 4.9cm 0.2cm 0.1cm 0cm]{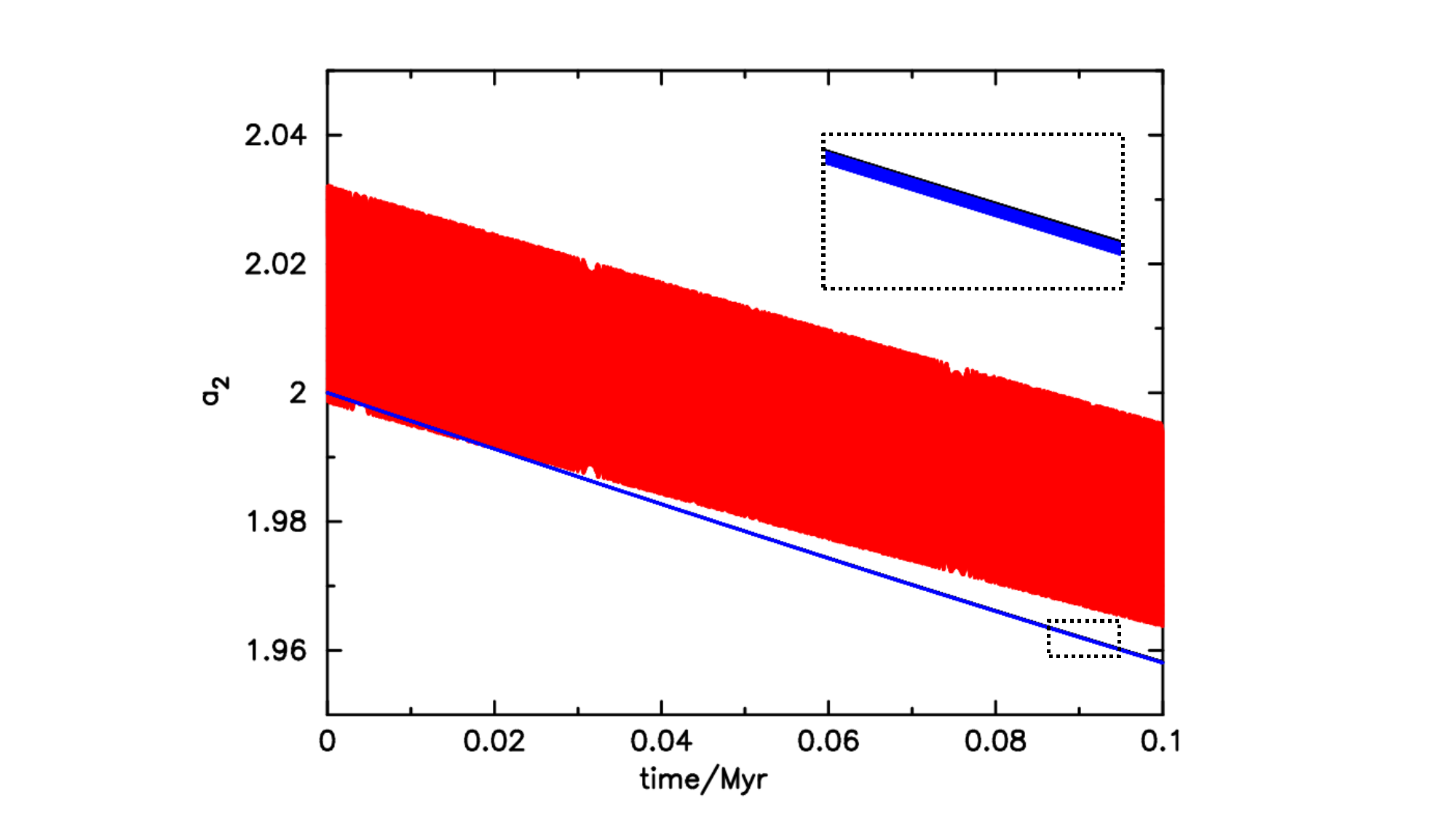}}
    \end{subfigure}
    \hfill 
    \begin{subfigure}[t]{0.48\textwidth}
        \raisebox{-\height}{\includegraphics[scale=0.29, angle=0, trim= 3.9cm 0cm 0.1cm 0cm]{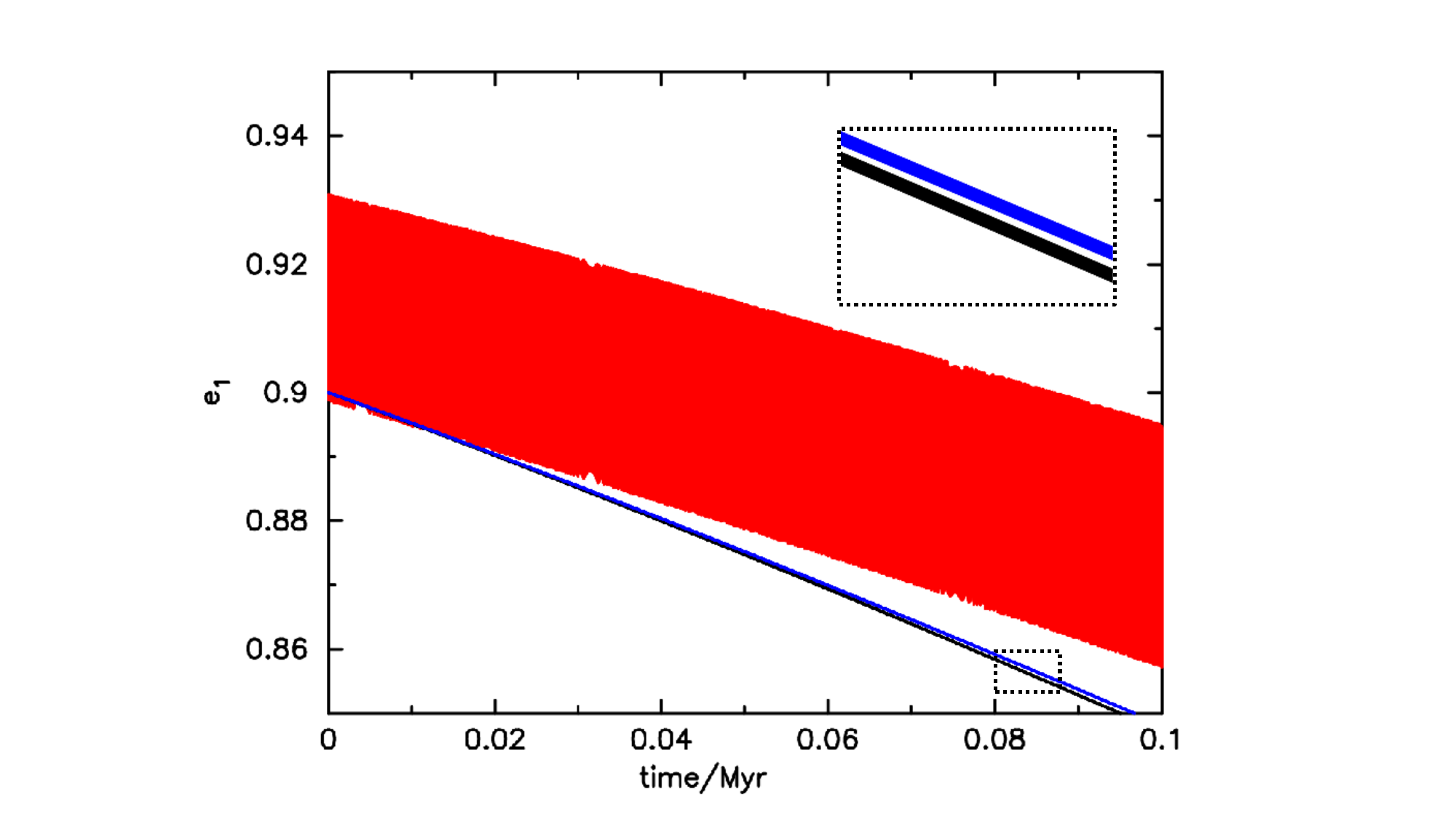}}
    \end{subfigure}
    \caption{Left panel: a comparison between the evolution of the outer orbit's semimajor axis, as predicted by a full tidal simulation (red), Equation \ref{de1dt} (black), and Equation \ref{de1dtJ} (blue). Right panel: comparison between the evolution of the inner orbit's eccentricity, as predicted by a full tidal simulation (red), Equation \ref{de1dt} (black), and Equation \ref{de1dtJ} (blue). In both cases, the blue and black lines almost coincide exactly with each other, although this is more true of the left panel than the right. A region in the lower right corner of each plot marked by the dashed rectangle has been magnified and displayed in the upper right corner to illustrate this.}
    \label{fig6}
\end{figure*}

Conservation of angular momentum manifests itself as 

\begin{equation}
0=\dot{J}_{\rm tot}=\dot{J}_{\rm 1}+\dot{J}_{\rm 2}+\dot{J}_{\rm spin},
\label{Jtot}
\end{equation}

\noindent where 

\begin{equation}
J_{\rm 1}^{\rm 2}=G\frac{m_{\rm 1}^{\rm 2}m_{\rm 2}^{\rm 2}}{m_{\rm 1}+m_{\rm 2}}a_{\rm 1}\left(1-e_{\rm 1}^{\rm 2}\right),
\label{J1}
\end{equation}

\begin{equation}
J_{\rm 2}^{\rm 2}=G\frac{\left(m_{\rm 1}+m_{\rm 2}\right)^{\rm 2}m_{\rm 3}^{\rm 2}}{m_{\rm 1}+m_{\rm 2}+m_{\rm 3}}a_{\rm 2},
\label{J2}
\end{equation}

\noindent and, as with rotational energy, we omit the effects of spin angular momentum and set $J_{\rm spin}=0$ for now ($\bm{J_{\rm spin}}$/$\bm{J_{\rm 2}}$=0.012). This leads to

\begin{equation}
\begin{split}
2a_{\rm 1}e_{\rm 1}\frac{ {\rm d}e_{\rm 1} } { {\rm d}t }=&\left(1-e_{\rm 1}^{\rm 2}\right)\frac{ {\rm d}a_{\rm 1} } { {\rm d}t }+ \\
&\sqrt{\frac{\left(m_{\rm 1}+m_{\rm 2}\right)^{\rm 3}m_{\rm 3}^{\rm 2}a_{\rm 1}\left(1-e_{\rm 1}^{\rm 2}\right)}{m_{\rm 1}^{\rm 2}m_{\rm 2}^{\rm 2}\left(m_{\rm 1}+m_{\rm 2}+m_{\rm 3}\right)a_{\rm 2}}}\frac{ {\rm d}a_{\rm 2} } { {\rm d}t },
\label{de1dt}
\end{split}
\end{equation}

\noindent where 

\begin{equation}
\frac{ {\rm d}a_{\rm 1} } { {\rm d}t }=\frac{Gm_{\rm 1}m_{\rm 2}}{2E_{\rm 1}^{\rm 2}}\frac{ {\rm d}E_{\rm 1} } { {\rm d}t },
\label{da1dt}
\end{equation}

\begin{equation}
\frac{ {\rm d}a_{\rm 2} } { {\rm d}t }=\frac{G\left(m_{\rm 1}+m_{\rm 2}\right)m_{\rm 3}}{2E_{\rm 2}^{\rm 2}}\frac{ {\rm d}E_{\rm 2} } { {\rm d}t },
\label{da2dt}
\end{equation}

\noindent which can be numerically integrated as an ordinary differential equation with $a_{\rm 1}$, $a_{\rm 2}$, and $e_{\rm 1}$ as integrated variables, with their first time derivative provided by Equations \ref{dE1dti}, \ref{dE2dti}, \ref{de1dt}, \ref{da1dt}, and \ref{da2dt}. The results of this integration indicate that setting $N=\frac{1}{2}$ fits the outer orbital shrinkage reasonably well (see Figure \ref{fig6}), and therefore we rewrite Equation \ref{dE2dti} as

\begin{equation}
\begin{split}
\frac{1}{E_{\rm 1}}\frac{ {\rm d}E_{\rm 2} } { {\rm d}t }&=C\frac{{\langle}r^{4.8}{\rangle}}{a_{\rm 1}^{4.8}}\left(\frac{e_{\rm 1}^{\rm 2}}{2}\right) \\
&\left(\frac{a_{\rm 1}}{0.2{\rm AU}}\right)^{4.8}\left(\frac{a_{\rm 2}}{2{\rm AU}}\right)^{-10.2}\frac{4q}{\left(1+q\right)^{2}} \\
&\left(\frac{R_{\rm 3}}{100{\rm R}_{\odot}}\right)^{5.2}\left(\frac{\tau}{0.534{\rm yrs}}\right)^{-1.0},
\label{dE2dt2}
\end{split}
\end{equation}

\noindent

\begin{figure}
\includegraphics[scale=0.29, angle=0, trim= 2.2cm 0cm 0.1cm 0cm]{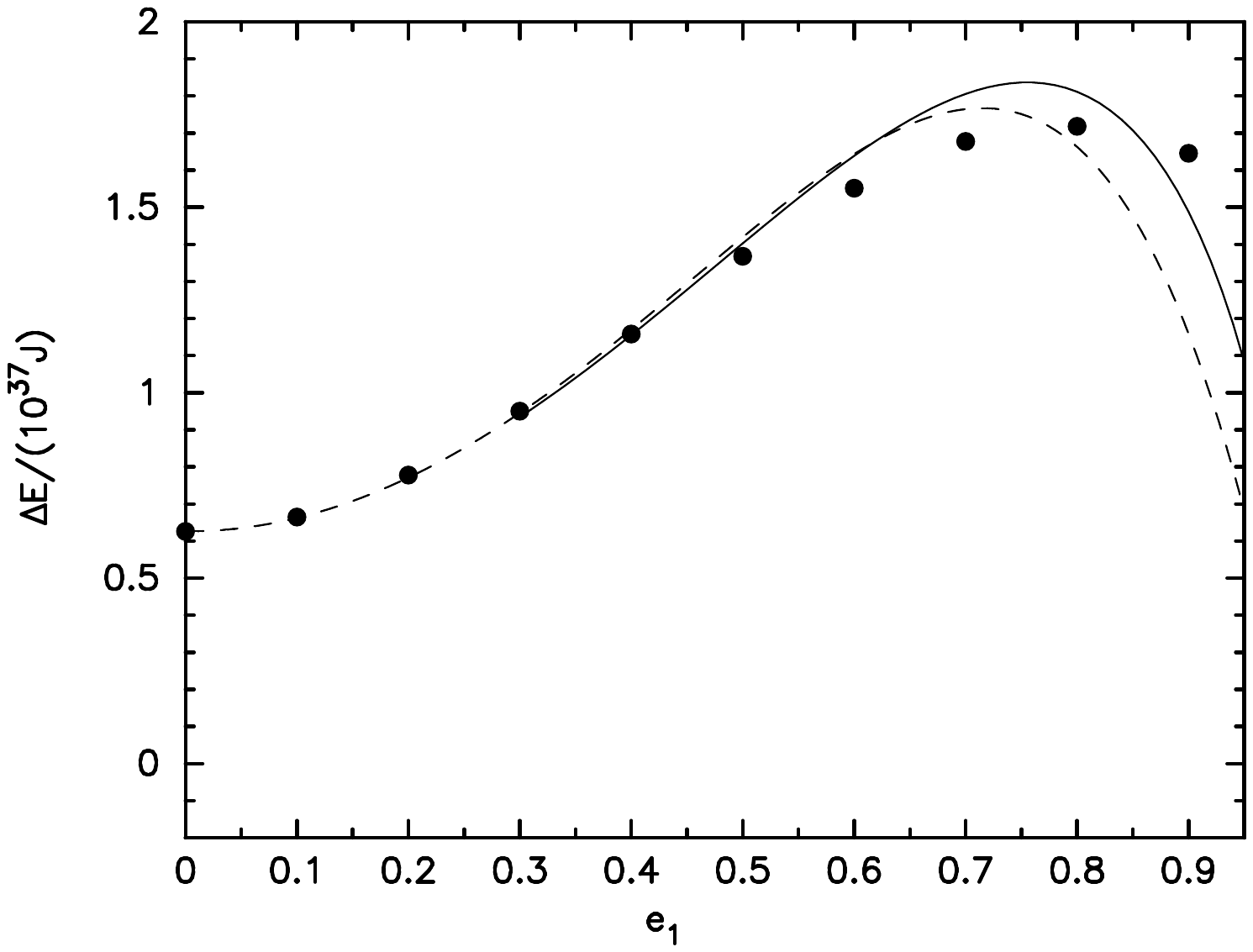}
\caption{Results of a numerical integration of Equations \ref{dE1dti}, \ref{dE2dt2}, \ref{de1dt}, \ref{da1dt}, and \ref{da2dt} overlaid on Figure \ref{fig2}. \label{fig7}}
\end{figure}

After re-performing this integration using Equations \ref{dE1dti}, \ref{dE2dt2}, \ref{de1dt}, \ref{da1dt}, and \ref{da2dt}, we present the results in Figure \ref{fig7}, where it can be seen that the lack of monotonicity is replicated.

But what about the spin energy and spin angular momentum we omitted earlier?

Including them analytically in the  calculations of Equations \ref{dE1dti}, \ref{dE2dt2}, \ref{de1dt} can be tricky, since it has already been established that, when TTs are present in a hierarchical triple system, the tertiary is not exactly tidally locked to the centre of mass of the inner binary \citep{2018MNRAS.479.3604G}. However, since the deviation from perfect spin-orbit resonance is small, we can estimate the influence of our omission by assuming perfect tidal locking, thereby roughly showing the magnitude of the energy and angular momentum absorbed by the tertiary's spin under that assumption. 

With tidal locking, the spin angular momentum of the tertiary is

\begin{equation}
J_{\rm spin}=I{\omega}=\frac{4}{35}m_{\rm 3}R_{\rm 3}^{\rm 2}\sqrt{\frac{G\left(m_{\rm 1}+m_{\rm 2}+m_{\rm 3}\right)}{a_{\rm 2}^{\rm 3}}},
\label{Jspin}
\end{equation}

\noindent where $I$ is the rotational inertia of the tertiary, and $\omega$ is the outer orbital angular velocity averaged over an orbital period. Here, we take advantage of the $I=\frac{4}{35}m_{\rm 3}R_{\rm 3}^{\rm 2}$ result for polytropes approximating red giants previously derived in \cite{2018MNRAS.479.3604G}. Hence, Equation \ref{de1dt} becomes

\begin{equation}
\begin{split}
2a_{\rm 1}e_{\rm 1}\frac{ {\rm d}e_{\rm 1} } { {\rm d}t }=&\left(1-e_{\rm 1}^{\rm 2}\right)\frac{ {\rm d}a_{\rm 1} } { {\rm d}t }+ \\
&\sqrt{\frac{\left(m_{\rm 1}+m_{\rm 2}\right)^{\rm 3}m_{\rm 3}^{\rm 2}a_{\rm 1}\left(1-e_{\rm 1}^{\rm 2}\right)}{m_{\rm 1}^{\rm 2}m_{\rm 2}^{\rm 2}\left(m_{\rm 1}+m_{\rm 2}+m_{\rm 3}\right)a_{\rm 2}}}\frac{ {\rm d}a_{\rm 2} } { {\rm d}t }-\\
&\frac{12}{35}\frac{m_{\rm 3}R^{\rm 2}_{\rm 3}}{a^{\rm 4}_{\rm 2}}\sqrt{\frac{m_{\rm 1}+m_{\rm 2}}{m_{\rm 1}m_{\rm 2}}}\sqrt{\frac{m_{\rm 1}+m_{\rm 2}+m_{\rm 3}}{m_{\rm 1}m_{\rm 2}}} \\
&\sqrt{a_{\rm 1}a^{\rm 3}_{\rm 2}\left(1-e_{\rm 1}^{\rm 2}\right)},
\label{de1dtJ}
\end{split}
\end{equation}

Finally, at the end of the calculation, the tertiary spin energy, under the assumption of perfect tidal locking, is

\begin{equation}
E_{\rm spin}=\frac{1}{2}I{\omega}=\frac{4}{35}m_{\rm 3}R_{\rm 3}^{\rm 2}\sqrt{\frac{G\left(m_{\rm 1}+m_{\rm 2}+m_{\rm 3}\right)}{a_{\rm 2}^{\rm 3}}}.
\label{Jspin}
\end{equation}

\begin{figure}
\includegraphics[scale=0.29, angle=0, trim= 2.2cm 0cm 0.1cm 0cm]{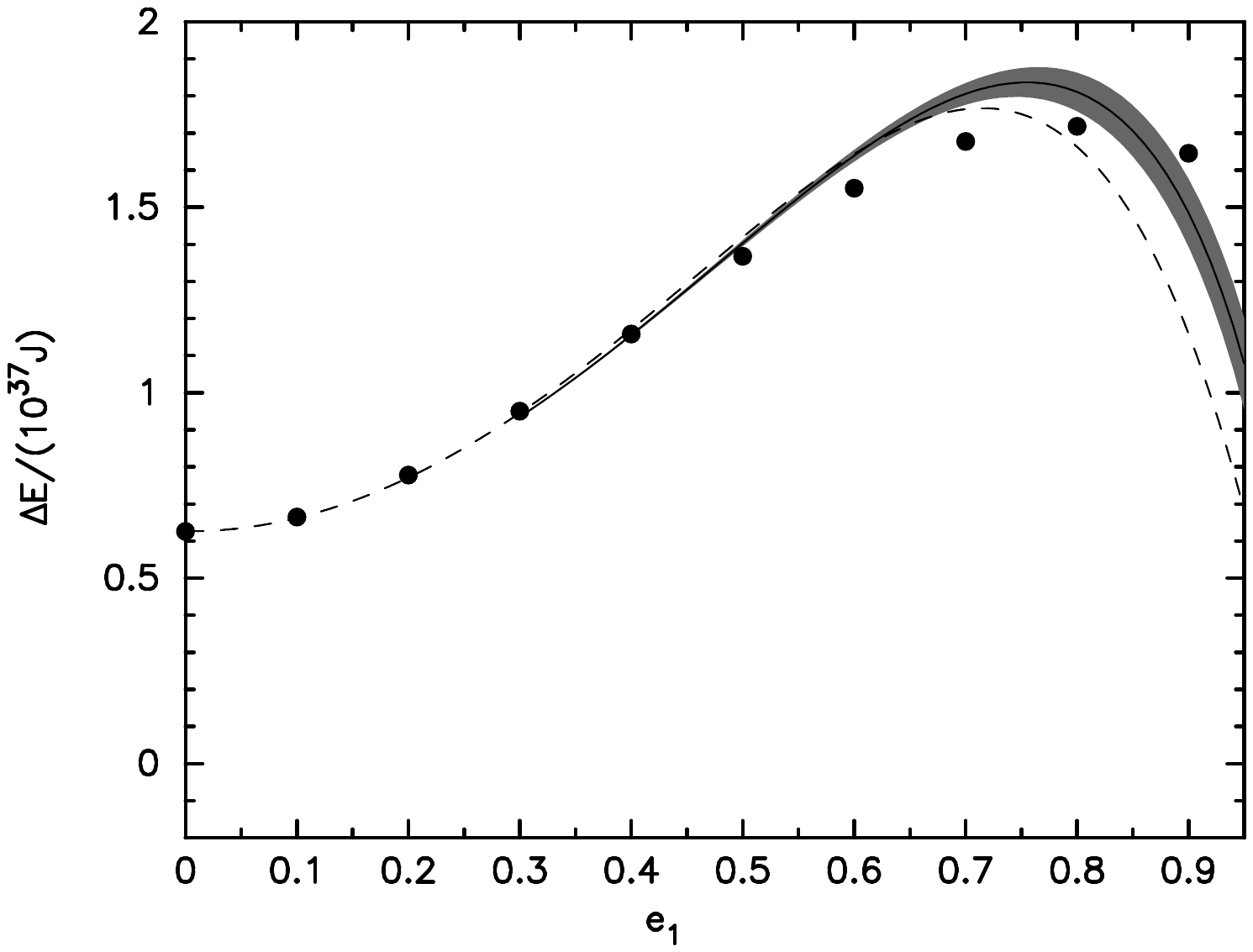}
\caption{Results of a numerical integration of Equations \ref{dE1dti}, \ref{dE2dt2}, \ref{da1dt}, and \ref{da2dt}, with the spin angular momentum of the tertiary accounted for via Equation \ref{de1dtJ}, overlaid on Figure \ref{fig2}. The grey area indicates the area that could be reached by adding or subtracting the final rotational energy of the tertiary, assuming perfect tidal synchronisation throughout. This provides an estimate of how much influence the tertiary's rotational angular momentum (which we are not able to analytically provide) and its rotational energy (which we omit from our calculations), could have had on the results. \label{fig8}}
\end{figure}

The final results of this set of calculations is shown in Figure \ref{fig8}, where it can be seen that the discrepancy is reconcilable with the energy and angular momentum carried by the tertiary to within an order of magnitude. This, coupled with the nonlinear effects that are present when adjusting the values of $a_{\rm 1}$, $a_{\rm 2}$, and $e_{\rm 1}$, not to mention the uncertainties caused by the characteristic oscillation in these parameters as they evolve, leads us to conclude that this omitted energy and angular momentum is consistent with the discrepancy seen. 

In summary, in cases where the inner binary is eccentric, Equation \ref{Eq1} ought to be replaced by the following:

\begin{equation}
\begin{split}
\frac{ {\rm d}E_{\rm 1} } { {\rm d}t }=&CE_{\rm 1}\frac{{\langle}r^{4.8}{\rangle}}{\left(0.2{\rm AU}\right)^{4.8}}\left(1-e_{\rm 1}^{\rm 2}\right)\left(\frac{a_{\rm 2}}{2{\rm AU}}\right)^{-10.2} \\
&\frac{4q}{\left(1+q\right)^{2}}\left(\frac{R_{\rm 3}}{100{\rm R}_{\odot}}\right)^{5.2}\left(\frac{\tau}{0.534{\rm yrs}}\right)^{-1.0},
\label{dE1dt}
\end{split}
\end{equation}

\begin{equation}
\begin{split}
\frac{ {\rm d}E_{\rm 2} } { {\rm d}t }=&CE_{\rm 1}\frac{{\langle}r^{4.8}{\rangle}}{\left(0.2{\rm AU}\right)^{4.8}}\left(\frac{e_{\rm 1}^{\rm 2}}{2}\right)\left(\frac{a_{\rm 2}}{2{\rm AU}}\right)^{-10.2} \\
&\frac{4q}{\left(1+q\right)^{2}}\left(\frac{R_{\rm 3}}{100{\rm R}_{\odot}}\right)^{5.2}\left(\frac{\tau}{0.534{\rm yrs}}\right)^{-1.0},
\label{dE2dt}
\end{split}
\end{equation}

\noindent and the evolution of $e_{\rm 1}$ follows conservation of angular momentum, determined by Equation \ref{de1dt}.

\subsection{Further Validation}

To validate the robustness of Equations \ref{dE1dt}, \ref{dE2dt}, and \ref{de1dt} when calculating for systems with initial conditions different to those of the Hypothetical Scenario configuration, we randomly choose a set of ($a_{\rm 1}$, $a_{\rm 2}$, $R_{\rm 3}$, $e_{\rm 1}$), and perform both the full tidal simulation and the calculation via Equations \ref{dE1dti}, \ref{dE2dt2}, \ref{de1dt}, \ref{da1dt}, and \ref{da2dt} for each set. The results of these simulations and calculations are shown in Table \ref{testconsq}, in terms of the amount of energy that is extracted from the inner orbit in $10^5$ years. The results for similar systems with circular orbits are also shown for comparison. The differences between the results are generally less than 10\%, with the errors being higher for higher eccentricities.

\begin{table*}
 \caption{Initial conditions and results of a set of arbitrarily selected test runs, in which $a_{\rm 1}$, $a_{\rm 2}$, $R_{\rm 3}$, and $e_{\rm 1}$ are varied, carried out to check the reliability of our empirical calculations (Equations \ref{dE1dti}, \ref{dE2dt2}, \ref{de1dt}, \ref{da1dt}, and \ref{da2dt}). The test runs have a duration of $10^5$ years, and the results are shown in the second of the final three columns in terms of the amount of energy that the inner binary loses over the course of the simulation. The corresponding results, calculated using our empirical equations, are presented in the first of the final three columns for comparison. The final column displays the results of the empirical equations of \protect\cite{2020MNRAS.491..264G}, assuming that the orbits are circular. The mutual errors between our empirical calculations and the direct simulations are on the order of 10\%.}
 \label{testconsq}
 \begin{tabular}{cccccccc}
  \hline
    Test \# & $a_{\rm 1}$/AU & $a_{\rm 2}$/AU & $R_{\rm 3}$/R$\odot$ & $e_{\rm 1}$ & Equation ${\Delta}E/10^{36}$J & Simulated ${\Delta}E/10^{36}$J & Circular ${\Delta}E/10^{36}$J \\
    \hline
    1 & 0.16 & 2.2 & 110 & 0.0 & $1.666$ & $1.548$ & $1.666$ \\
    2 & 0.17 & 2.4 & 120 & 0.1 & $1.436$ & $1.348$ & $1.358$ \\
    3 & 0.18 & 2.6 & 130 & 0.2 & $1.393$ & $1.314$ & $1.131$ \\
    4 & 0.19 & 2.8 & 140 & 0.3 & $1.448$ & $1.362$ & $0.959$ \\
    5 & 0.20 & 3.0 & 150 & 0.4 & $1.544$ & $1.434$ & $0.825$ \\
    6 & 0.21 & 3.2 & 160 & 0.5 & $1.632$ & $1.490$ & $0.719$ \\
    7 & 0.22 & 3.4 & 170 & 0.6 & $1.669$ & $1.502$ & $0.634$ \\
    8 & 0.23 & 3.6 & 180 & 0.7 & $1.602$ & $1.449$ & $0.564$ \\
    9 & 0.24 & 3.8 & 190 & 0.8 & $1.370$ & $1.319$ & $0.506$ \\
    10 & 0.25 & 4.0 & 200 & 0.9 & $0.888$ & $1.100$ & $0.457$ \\
  \hline
 \end{tabular}
\end{table*}

    

However, we caution that due to the omission of tertiary angular momentum and spin energy, our approximation equations break down when $q$ deviates significantly from 1. To illustrate this point, we repeat the validation process above with random values of $q$ added to the mix, the results of which are shown in Table \ref{testparams2}, where it can be seen that, at some values of $q$, the errors involved could be at least as large as 300\%. The reason for this is that, for values of $q$ that deviate significantly from 1, an injection of the same amount of angular momentum will cause a much larger variation of eccentricity. This, in turn, will lead to a nonlinear deviation which will increase the errors as time progresses.

\begin{table*}
 \caption{Similar to Table \ref{testconsq}, but the number of varied parameters was expanded relative to that tested in Table \ref{testconsq}, and now includes $m_{\rm 1}$ and $m_{\rm 2}$, and subsequently their mass ratio $q$. It can be seen that our empirical equations perform worse than in the $q=1$ case, due to the inner binary eccentricity being much more sensitive to angular momentum variations when $q$ deviates significantly from 1.}
 \label{testparams2}
 \begin{tabular}{ccccccccc}
  \hline
    $m_{\rm 1}$/M$\odot$ & $m_{\rm 2}$/M$\odot$ & $q(=m_{\rm 1}/m_{\rm 2}$) & $a_{\rm 1}$/AU & $a_{\rm 2}$/AU & $R_{\rm 3}$/R$\odot$ & $e_{\rm 1}$ & Equation ${\Delta}E/10^{36}$J & Simulated ${\Delta}E/10^{36}$J \\
    \hline
    0.15 & 0.99 & 0.152 & 0.16 & 2.2 & 110 & 0.0 & 0.177 & 0.165 \\
    0.30 & 0.98 & 0.306 & 0.17 & 2.4 & 120 & 0.1 & 0.675 & 0.444 \\
    0.45 & 0.97 & 0.464 & 0.18 & 2.6 & 130 & 0.2 & 1.193 & 0.775 \\
    0.60 & 0.96 & 0.625 & 0.19 & 2.8 & 140 & 0.3 & 1.684 & 1.161 \\
    0.75 & 0.95 & 0.789 & 0.20 & 3.0 & 150 & 0.4 & 2.058 & 1.574 \\
    0.90 & 0.94 & 0.957 & 0.21 & 3.2 & 160 & 0.5 & 2.250 & 1.964 \\
    1.05 & 0.93 & 1.129 & 0.22 & 3.4 & 170 & 0.6 & 2.224 & 2.285 \\
    1.20 & 0.92 & 1.304 & 0.23 & 3.6 & 180 & 0.7 & 1.976 & 2.486 \\
    1.35 & 0.91 & 1.484 & 0.24 & 3.8 & 190 & 0.8 & 1.517 & 2.630 \\
    1.50 & 0.90 & 1.667 & 0.25 & 4.0 & 200 & 0.9 & 0.859 & 3.319 \\
  \hline
 \end{tabular}
\end{table*}

\begin{figure}
\includegraphics[scale=0.29, angle=0, trim= 2.2cm 0cm 0.1cm 0cm]{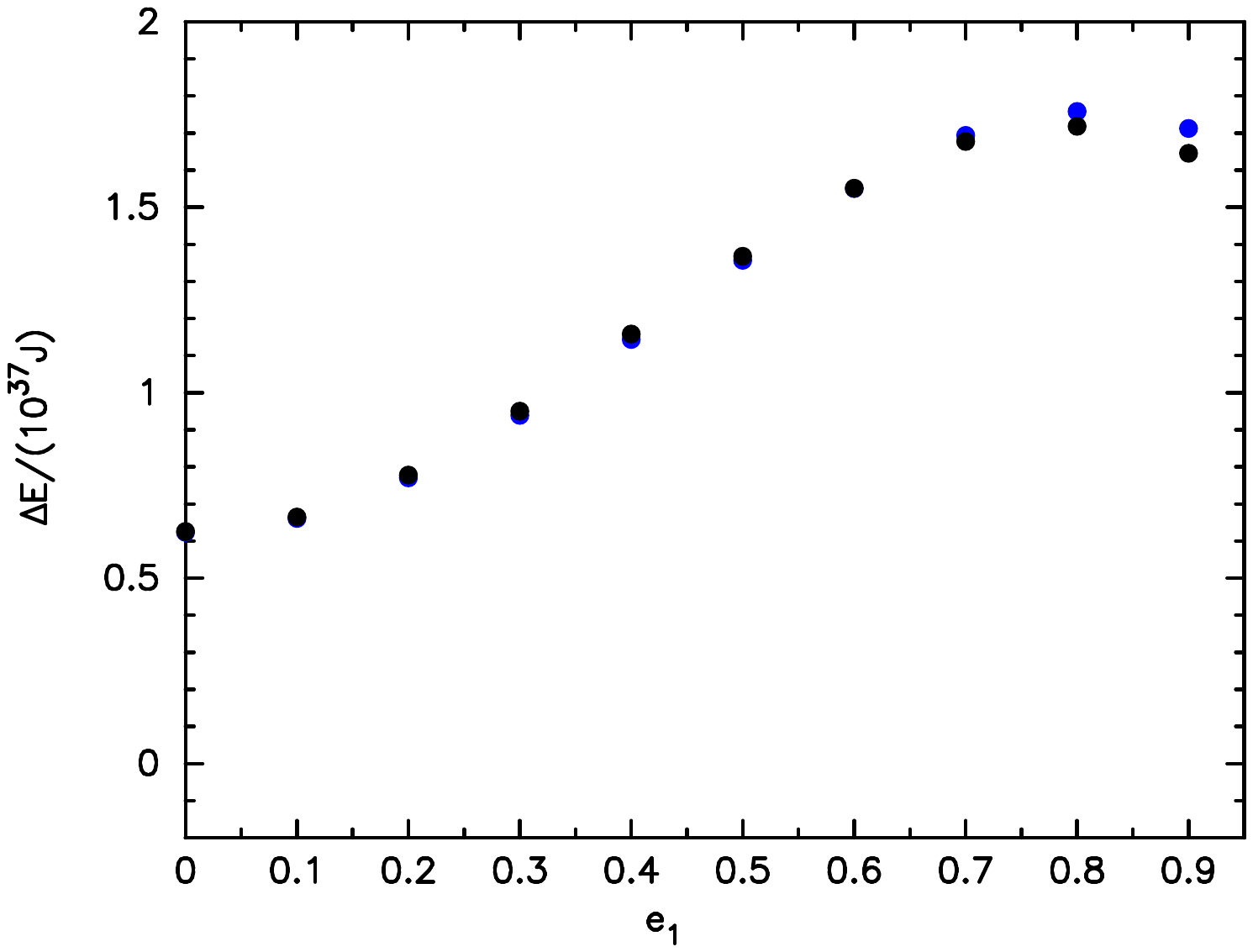}
\caption{Tidal simulation results of Figure \ref{fig2} (black points), compared to that of the same integration, except the initial orientation of the inner binary's semimajor axis is rotated 60 degrees (blue points). The disparity shows how sensitive the results are at high inner binary eccentricities. \label{fig9}}
\end{figure}

So just how bad is this nonlinearity? To quantify this, we repeat the full tidal simulations that led to Figure \ref{fig2}, except this time, we modify the argument of pericentre by 60 degrees. If this nonlinearity did not exist, then doing so should only be equivalent to starting the simulation 1/6th of an outer orbit (roughly 0.25 years) earlier - a non-issue. However, we find noticeably different results for a $10^5$ year run (see Figure \ref{fig9}) beyond $e_{\rm 1}=0.7$, indicating that this is roughly the eccentricity at which these effects become significant.

\subsection{Retrograde Orbits}

\begin{figure}
\includegraphics[scale=0.29, angle=0, trim= 2.2cm 0cm 0.1cm 0cm]{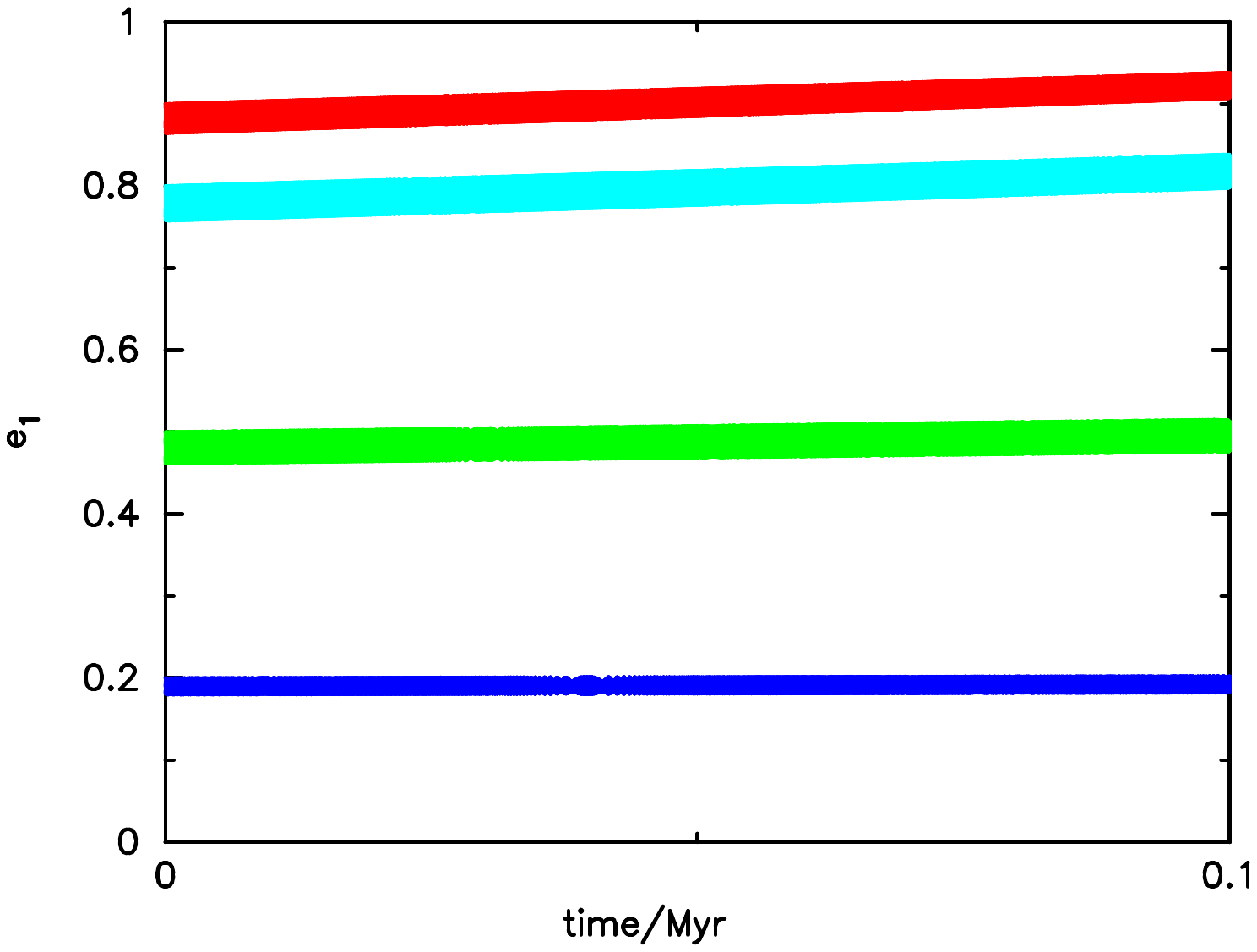}
\caption{Evolution of inner binary eccentricity as a function of time in a retrograde orbit. The different colours indicate different initial values of $e_{\rm 1}$: blue - 0.2, green - 0.5, cyan - 0.8, red - 0.9. In these cases, the eccentricities increase to compensate for the angular momentum loss of the outer orbit. This can lead to instabilities. \label{fig10}}
\end{figure}

Finally, we relax the condition that all orbits are prograde. We find that, for retrograde orbits where all the other orbital parameters are identical to our Hypothetical Scenario, the outer orbit also shrinks at similar rates to that of the prograde case. This is to be expected - after all, if our explanation of the outer orbital shrinkage is correct, then as long as the angular velocity of the tertiary in its orbit is out of sync with that of the apsidal precession of the inner binary, the quadrupole of the averaged inner orbit should always extract energy from the outer orbit. We also find that the inner binary eccentricity often increases in such configurations (Figure \ref{fig10}). This is also to be expected, as the angular momentum of the outer orbit decreases much faster than that of the inner orbit for even moderately high eccentricities, and the only way to balance this decrease is for the inner binary eccentricity to increase.

In many cases, this increase of the inner binary eccentricity destabilises the triple. This is a natural result of the eccentricity increasing to 1, and thereafter no longer being able to compensate for the evolution of angular momentum, ultimately unbinding the inner binary. It is for this reason that a retrograde version of Table \ref{testparams2} is not possible.


\section{Discussion}

Throughout this paper, we adopt a very specific viscoelastic model with a ${\tau}$ value of 0.534 years, which may require 
some justification, other than the fact that this was the value that we found in our earlier work on this system. 

A viscoelastic model was chosen, since TTs usually involve fast tides, and viscoelastic models tend to model fast tides well. The 
elastic components of the viscoelastic model were set to zero, since any elastic response of the red giant envelope (which can be significant where tidal amplitudes are concerned) contributes relatively little to tidal dissipation. TTs are special, in that they manifest as fast tides in giant stars. Giant stars rarely 
undergo fast tides in the binary case, as their large radius results in long orbits, and hence low tidal forcing frequencies. 
With TTs, however, the forcing frequency is dictated by the orbital period of the inner binary, allowing for much higher forcing 
frequencies than what can be allowed by an orbit around the tertiary. This was the primary motivation behind our adoption of a 
non-elastic viscoelastic model.

Regarding the value of ${\tau}$ which was used throughout this paper, we perform a sanity check by comparing our outer orbit 
circularisation results in Figure \ref{fig1} with those of \cite{1989A&A...220..112Z}. Using their Equations 15 and 21, and 
assuming that our inner binary can be treated as a single star of mass $m_{\rm 1}+m_{\rm 2}$, we arrive at an outer orbit 
circularisation timescale $t_{circ}$ of 222 years, or ${\log}(t_{circ}/Myr)$=-3.65. This is in excellent agreement with our 
Figure \ref{fig1}. Before we conclude that a ${\tau}$ value of 0.534 years is reasonable, however, we note that, in our original 
2018 paper, we used a fast tide dissipation timescale prescription consistent with \cite{2005ASPC..333....4Z} (see also \citealt{2007ApJ...655.1166P} and \citealt{2009ApJ...704..930P}) instead of 
\cite{1977ApJ...211..934G}, the latter of which seems to be more favoured by more recent studies 
\citep[e.g.][]{2012MNRAS.422.1975O,2020MNRAS.491..923D,2020MNRAS.497.3400D,2020ApJ...888L..31V}. Recalibrating ${\tau}$ using 
the \cite{1977ApJ...211..934G} fast tides prescription yields ${\tau}=$7 years, slowing down all the effects of TTs in this paper 
by 1.1 dex. However, we caution that TTs might not behave the way tides are modelled in these more advanced simulations. For instance, in a few extreme cases where the tidal perturbation frequencies are high, \cite{2012MNRAS.422.1975O} find ``negative effective viscosity" effects, which should see the inner 
orbital separation be increased by TTs. No value of ${\tau}$ yields such a result, since no matter how the tertiary radius expands or contracts, TTs will always 
result in the orbital energy of the inner binary being dumped onto the tertiary. The issue of 
exactly what fast tides prescription is most suitable for TTs will ultimately be determined observationally, such as by 
observations of TIC 242132789 \citep{2023MNRAS.521.2114G}.

Our results for tertiary tides in a hierarchical triple system in which the outer orbit is eccentric are straightforward: the tides act like two-body tides, circularising the outer orbit while retaining its angular momentum, if ${\tau}$ is small enough. The large radius of the tertiary will guarantee that this process occurs on very short timescales, after which tertiary tides will act according to Equation \ref{Eq1}. While simple, this result has an important implication, in that systems undergoing tertiary Roche Lobe overflow from a giant tertiary, such as \citep[e.g.][]{2014MNRAS.438.1909D,2019ApJ...876L..33P,2020MNRAS.496.1819L,2023MNRAS.518..526G}, as well as progenitor systems of many tertiary-donated circumtriple common envelope systems (\citealt{2021MNRAS.500.1921G,2020MNRAS.498.2957C}, see also \citealt{2020MNRAS.493.1855D}), are likely to begin their mass transfer process with a circularised outer orbit. This should greatly simplify future calculations for such systems.

As for the effects of TTs in systems where the inner binary is eccentric, we find that it shrinks both the inner and outer orbits, while its effect on the inner binary eccentricity depends on the initial configuration:
\begin{itemize}
  \item For prograde orbits, the only way that angular momentum can be conserved when both the inner and outer orbits shrink is by decreasing the inner binary eccentricity.
  \item For retrograde orbits, the direction in which the inner binary eccentricity evolves is determined by which orbit, inner or outer, loses angular momentum due to orbital shrinkage faster. In most cases where there is moderately high inner binary eccentricity, the outer orbit will be much more prone to losing angular momentum when a similar amount of orbital energy is extracted from both orbits, and therefore the inner binary eccentricity will increase.
  \item When the inner binary eccentricity does increase in the retrograde case, it eventually leads to a runaway effect in which the outer orbital shrinkage becomes dominant, and the balance of angular momentum can subsequently only be maintained by further increasing the inner binary eccentricity, which in turn further reinforces the outer orbital shrinkage. If TTs act for long enough, this will eventually disrupt the system. 
\end{itemize}

Omitting the variations of the tertiary's spin, we find that the equations \ref{dE1dt}, \ref{dE2dt}, and \ref{de1dt} approximate the evolution well. Since the outer orbit shrinks for an eccentric inner orbit, this is expected to have an effect on the evolution of the host hierarchical triple, which is even more pronounced than those found in circular orbits. As the outer orbit shrinks, the total amount of energy dissipated per unit time increases drastically for both orbits, and since the timescale of inner orbital eccentricity evolution is similar to that at which both the inner and outer semimajor axes shrink, this could potentially lead to extremely tight inner binaries at the end of the tertiary's giant phase. On the other hand, at extremely high eccentricities, it is possible that the outer orbit may shrink much faster than the inner orbit, leading to a dynamically unstable triple \citep[e.g.][]{2001MNRAS.321..398M,2018MNRAS.474...20H}. One potential way by which such eccentricities can be injected into hierarchical triples is via galactic tides \citep{2024ApJ...972L..19S}. However, the efficacy of this process in generating close hierarchical triples, and the prevalence of such high-inner-binary-eccentricity hierarchical triples remains to be seen.


In terms of the implications of eccentric TTs on stellar evolution in general, tertiary mass transfer aside, it should also be noted that the retrograde case could potentialy revive many formation channels for exotic objects once thought to be unrealistic. For instance, collisional SNe Ia \citep{2015MNRAS.454L..61D} were once thought to be impossibly rare, as the eccentricity required for a fine-tuned head-on collision between a WD pair is extreme. However, with a retrograde tertiary close enough to provide TTs, a runaway increase of the eccentricity is now possible. { No corresponding process that induces such a runaway increase can be found in classical two-body tides \citep[e.g.][]{1981A&A....99..126H}}.

Regarding gravitational wave sources, it should be pointed out that, should TTs serve to drive a black hole binary closer together, the efficiency at which it does so could be enhanced by several orders of magnitude in the eccentric case. This is because the rate at which energy is extracted from the inner binary increases as $a^{-10.2}_{\rm 2}$, therefore any shrinkage of the outer orbit, such as that provided by inner binary eccentricity, can easily reduce the time required to merge the pair.

As with the coplanar case, in order to trigger TTs, the tertiary of the triple system must have a radius that comes close to filling its Roche Lobe. In this paper, we study the case in which the tertiary is a red giant; as such, our results should apply to cases where the tertiary has an envelope conducive to tidal dissipation, such as a convective envelope. Thus, our equations apply to hierarchical triples with tertiaries that are either low-mass main sequence stars, giants, or rapidly spinning high-mass main sequence stars that are undergoing Eddington-Sweet circulation. If this condition is not satisfied, such as when the tertiary has a radiative envelope, TTs are expected to be much weaker. However, since the physics of how tides act within such systems is not well known, exactly how much weaker would be difficult to quantify.

As previously mentioned, the work presented in this paper forms the basis of an investigation into tertiary tides in non-coplanar hierarchical triples, where other tertiary effects, such as Lidov-Kozai cycles \citep{1910AN....183..345V,1962P&SS....9..719L,1962AJ.....67..591K}, will drive up the inner eccentricity from zero. All of this, however, will be the contents of our next paper.

\section*{Acknowledgements}


YG was a Royal Society K.C. Wong International Fellow at the time this paper was started, and as such acknowledges funding from the Royal Society and the K.C. Wong Education Foundation.

\noindent TB is supported by an appointment to the NASA Postdoctoral Program at the NASA Ames Research Center, administered by Oak Ridge Associated Universities under contract with NASA.

\noindent DP acknowledges the ASPIRE program for providing the opportunity to participate in this project.

\noindent ST acknowledges support from the Netherlands Research Council NWO (VENI 639.041.645 and VIDI 203.061 grants).

\noindent Many thanks to Thomas Baycroft and Cl\'{e}ment Bonnerot for useful discussions

\section*{Data Availability Statement}

The data underlying this article will be shared on reasonable request to the corresponding author.








\appendix

\section{Viscoelastic Model}
\label{appA}

The following is a summary of the viscoelastic model from \cite{2018MNRAS.479.3604G}, which we also adopt for the simulations in the present paper.

In our hierarchical triple system consisting of three stars with masses of $m_{\rm 1}$, $m_{\rm 2}$ and $m_{\rm 3}$, the inner binary pair are treated as point masses, while the tertiary is modelled as an ellipsoid, the mass distribution of which is described by four parameters: its mean radius $R_3$, and the gravity field coefficients $J_2$, $C_{22}$ and $S_{22}$. The Cartesian coordinate system (${\bm I}$,${\bm J}$,${\bm K}$) herein established for the purpose of defining $J_2$, $C_{22}$ and $S_{22}$ has its point of origin at the centre of mass of the tertiary, and ${\bm K}$ aligns with its axis of maximum inertia, perpendicular to the orbital plane. Thus, neglecting terms of order $(R_{\rm 3}/r)^3$ and higher, the gravitational potential of the tertiary becomes

\begin{equation}
\begin{split}
V ({\bm r}) =& - \frac{G m_3}{r} - \frac{G m_3 R_{\rm 3}^2 J_2}{2 r^3} \\
& - \frac{3 G m_3 R_{\rm 3}^2}{r^3}  \big(  C_{22} \cos {2 \gamma}  - S_{22} \sin {2 \gamma} \big), 
\label{c1} 
\end{split}
\end{equation}
\noindent where
\begin{equation}
\begin{split}
&\cos {2 \gamma} =  ({\bm I} \cdot {\bm {\hat r}})^2 - ({\bm J} \cdot {\bm {\hat r}})^2, \\
&\sin {2 \gamma} = - 2 ({\bm I} \cdot {\bm {\hat r}}) ({\bm J} \cdot {\bm {\hat r}}) \ ,
\end{split}
\end{equation}
\noindent and ${\bm r}$ is any generic position under the coordinate system (${\bm I}$,${\bm J}$,${\bm K}$), while ${\bm {\hat r}} = {\bm r} / r $ is the corresponding unit vector. ${\gamma}$ is defined as ${\gamma} = \theta - f$, where $\theta$ is the rotation angle, and ${f}$ is the true longitude.

Likewise, the total potential energy of the system becomes
\begin{equation}
U ({\bm r}_1,{\bm r}_2) = - \frac{G m_1 m_2}{| {\bm r}_2-{\bm r}_1 |} + m_1 V({\bm r}_1) + m_2 V({\bm r}_2) \ , \label{c3} 
\end{equation}
where ${\bm r}_i = {\bm R}_i - {\bm R}_3$, and ${\bm R}_i$ is the position of the star with mass $m_i$ in an inertial frame.

Thus, the equations of motion in an inertial frame are
\begin{equation}
\frac{d^2 {\bm R}_i}{dt^2} = - \frac{1}{m_i} \frac{\partial U}{\partial {\bm R}_i} = - \frac{1}{m_i} \frac{\partial U}{\partial {\bm r}_i} \ , \label{161223b} 
\end{equation}
\begin{equation}
\frac{d^2 {\bm R}_3}{dt^2} = - \frac{1}{m_3} \frac{\partial U}{\partial {\bm R}_3} = \frac{1}{m_3} \left( \frac{\partial U}{\partial {\bm r}_1} + \frac{\partial U}{\partial {\bm r}_2} \right) \ , \label{161223c} 
\end{equation}
where
\begin{equation}
\begin{split}
&\frac{\partial U}{\partial {\bm r}_i} = (-1)^i \frac{G m_1 m_2}{|{\bm r}_2-{\bm r}_1|^3} ({\bm r}_2-{\bm r}_1) 
+ \frac{G m_i m_3}{r_i^3} {\bm r}_i  \\ 
&+ \frac{3 G m_i m_3 R_{\rm 3}^2}{2 r_i^5} \left[ J_2 + 6 \big(  C_{22} \cos {2 \gamma}_i  - S_{22} \sin {2 \gamma}_i \big) \right]  {\bm r}_i  \\ 
&- \frac{6 G m_i m_3 R_{\rm 3}^2}{r_i^5}  \big( C_{22} \sin {2 \gamma}_i  + S_{22} \cos {2 \gamma}_i \big) {\bm K} \times {\bm r}_i \ . 
\label{161223d} 
\end{split}
\end{equation}

\noindent while in an astrocentric frame, they become 
\begin{equation}
\frac{d^2 {\bm r}_i}{dt^2} = - \left( \frac{1}{m_i} + \frac{1}{m_3} \right) \frac{\partial U}{\partial {\bm r}_i} - \frac{1}{m_3} \frac{\partial U}{\partial {\bm r}_j}  \ , \label{161223e}
\end{equation}
where $i = 1,2$ and $j=3-i$.

The tertiary experiences a torque
\begin{equation}
I_{\rm 3} \frac{d {\Omega}}{dt} = \left( {\bm r}_1 \times \frac{\partial U}{\partial {\bm r}_1} + {\bm r}_2 \times \frac{\partial U}{\partial {\bm r}_2} \right) \cdot {\bm K}  \ , \label{161223f} 
\end{equation}
where $I_{\rm 3}$ is the principal moment of inertia of $m_{\rm 3}$ along ${\bm K}$, and hence the rotation angle $\dot \theta = {\Omega}$ can be expressed as
\begin{equation}
\begin{split}
\frac{d^2 \theta}{dt^2} &= -\frac{6Gm_1m_3R_{\rm 3}^2}{I_{\rm 3}r_1^3} \left[ C_{22}\sin2{\gamma}_1 + S_{22}\cos2{\gamma}_1\right] \\  
&-\frac{6Gm_2m_3R_{\rm 3}^2}{I_{\rm 3}r_2^3} \left[ C_{22}\sin2{\gamma}_2 + S_{22}\cos2{\gamma}_2\right] \ .
\label{161223g} 
\end{split}
\end{equation}

Since the tertiary's radius is assumed to stay constant throughout the simulations in this paper, our prescriptions for time-variant tertiary radius will be omitted here.

To account for the effects of tidal deformation, we apply a Maxwell model:

\begin{equation}
\begin{split}
&J_2+\tau\dot{J}_2 =  J_2^{r} + J_2^t \ ,\\
&C_{22}+\tau\dot{C}_{22}  =  C_{22}^t \ ,\\
&S_{22}+\tau\dot{S}_{22} = S_{22}^t \ ,
\label{max1}
\end{split}
\end{equation}
where
\begin{equation}\label{j2r}
J_2^{r} = k_2 \frac{{\Omega}^2R_{\rm 3}^3}{3Gm_3},
\end{equation}
and 
\begin{equation}
\begin{split}
&J_2^t=k_2 \frac{m_1}{2m_3}\left(\frac{R_{\rm 3}}{r_1}\right)^3 + k_2 \frac{m_2}{2m_3}\left(\frac{R_{\rm 3}}{r_2}\right)^3 \ ,\\
&C_{22}^t=\frac{k_2}{4}\frac{m_1}{m_3}\left(\frac{R_{\rm 3}}{r_1}\right)^3\cos2{\gamma}_1 + \frac{k_2}{4}\frac{m_2}{m_3}\left(\frac{R_{\rm 3}}{r_2}\right)^3\cos2{\gamma}_2 \ , \\
&S_{22}^t=-\frac{k_2}{4}\frac{m_1}{m_3}\left(\frac{R_{\rm 3}}{r_1}\right)^3\sin2{\gamma}_1 -\frac{k_2}{4}\frac{m_2}{m_3}\left(\frac{R_{\rm 3}}{r_2}\right)^3\sin2{\gamma}_2 \ .
\label{max2}
\end{split}
\end{equation}

Here, $\tau = \tau_v + \tau_e$, where $\tau_v$ and $\tau_e$ are the viscous and elastic relaxation times, respectively (see \citealt{2014A&A...571A..50C}). For our original calibration resulting in $\tau = $0.534 yr, we assumed $\tau_e=0$, since the elastic component is not expected to be significant within a red giant.

\section{Relation between Quadrupole Moment of an Averaged Keplerian Orbit and its Eccentricity}
\label{appB}

For an eccentric Keplerian orbit, the mass distribution averaged over one orbit has a quadrupole moment. For any given hierarchical triple, the magnitude of this quadrupole moment for its inner binary is directly related to the influence of tertiary tides on its outer orbit. Here, we provide a calculation of how the strength of this quadrupole moment is related to its eccentricity $e$ .

Establish a set of Cartesian coordinates with the point of origin at the centre of mass of the inner binary, with the x-axis coinciding with the major axis of the inner binary orbit. Due to symmetry, the $3{\times}3$ quadrupole tensor $\mathbf{Q}_{ij}$ only has non-zero terms on its diagonal components. Since the trace of $\mathbf{Q}_{ij}$ is zero, that leaves only two components that are independent.

These two components are

\begin{equation}
\begin{split}
Q_{xx}=&\frac{q}{\left(1+q\right)^{2}} \left(m_{\rm 1}+m_{\rm 2}\right) \\
&\frac{1}{T}{\int}_{0}^{T} \left[3\left(r{\cos}f\right)^2-r^2\right] {\rm d}t, \\
Q_{yy}=&\frac{q}{\left(1+q\right)^{2}} \left(m_{\rm 1}+m_{\rm 2}\right) \\
&\frac{1}{T}{\int}_{0}^{T} \left[3\left(r{\sin}f\right)^2-r^2\right] {\rm d}t, 
\label{appen1}
\end{split}
\end{equation}

\noindent where $f$ is the true anomaly, the integration sign integrates over a single orbit, and all the other symbols are the same as those used throughout the paper.

The strength of this quadrupole moment, where TTs are concerned, is proportional to the amplitude of the gravitational field variation in the direction of the tertiary, and is simply the difference between the two:

\begin{equation}
\begin{split}
Q_{xx}-Q_{yy}=&\frac{q}{\left(1+q\right)^{2}} \left(m_{\rm 1}+m_{\rm 2}\right) \\
&\frac{1}{T}{\int}_{0}^{T} \left[3\left(r{\cos}f\right)^2-3\left(r{\sin}f\right)^2\right] {\rm d}t \\
=&\frac{q}{\left(1+q\right)^{2}} \left(m_{\rm 1}+m_{\rm 2}\right) \\
&\frac{3}{T}{\int}_{0}^{T} \left[2\left(r{\cos}f\right)^2-r^2\right] {\rm d}t.
\label{appen2}
\end{split}
\end{equation}

Again adopting the methods of \cite{2017ApJ...837L...1M}, we note that 

\begin{equation}
\frac{1}{T}{\int}_{0}^{T} r {\rm d}t=a_{\rm 1}\left(1+\frac{e_{\rm 1}^{2}}{2}\right),
\label{appen3}
\end{equation}

\begin{equation}
\frac{1}{T}{\int}_{0}^{T} r^{2} {\rm d}t=a^{2}_{\rm 1}\left(1+\frac{3}{2}e_{\rm 1}^{2}\right),
\label{appen4}
\end{equation}

\noindent and also that

\begin{equation}
r=\frac{a_{\rm 1}\left(1-e_{\rm 1}^{2}\right)}{1+e_{\rm 1}{\cos}f},
\label{appen5}
\end{equation}

\noindent therefore

\begin{equation}
\begin{split}
&\frac{3}{T}{\int}_{0}^{T} \left[2\left(r{\cos}f\right)^2-r^2\right] {\rm d}t \\
=&\frac{3}{T}{\int}_{0}^{T} \left[2\left(\frac{a_{\rm 1}\left(1-e_{\rm 1}^{2}\right)-r}{e_{\rm 1}}\right)^2-r^2\right] {\rm d}t \\
=&\frac{3}{T}{\int}_{0}^{T} \left[\frac{2}{e_{\rm 1}^{2}}\left[a_{\rm 1}\left(1-e_{\rm 1}^{2}\right)-r\right]^{2}-r^2\right] {\rm d}t \\
=&\frac{6}{Te_{\rm 1}^{2}}\int_{0}^{T} \left[a_{\rm 1}\left(1-e_{\rm 1}^{2}\right)-r\right]^{2} {\rm d}t-3\frac{1}{T}{\int}_{0}^{T} r^{2} {\rm d}t \\
=&\frac{6}{e_{\rm 1}^{2}}\left[ a^{2}_{\rm 1}\left(1-e_{\rm 1}^{2}\right)^{2}-2a_{\rm 1}\left(1-e_{\rm 1}^{2}\right)\frac{1}{T}{\int}_{0}^{T}r{\rm d}t+\right. \\
&\left.\frac{1}{T}{\int}_{0}^{T} r^{2} {\rm d}t \right] -3\frac{1}{T}{\int}_{0}^{T} r^{2} {\rm d}t \\
=&\frac{6}{e_{\rm 1}^{2}}\left[ a^{2}_{\rm 1}\left(1-e_{\rm 1}^{2}\right)^{2}-2a^{2}_{\rm 1}\left(1-e_{\rm 1}^{2}\right)\left(1+\frac{e_{\rm 1}^{2}}{2}\right)+\right. \\
&\left.a^{2}_{\rm 1}\left(1+\frac{3}{2}e_{\rm 1} ^{2}\right)\right]     -3a^{2}_{\rm 1}\left(1+\frac{3}{2}e_{\rm 1}^{2}\right) \\
=&\frac{15}{2}a^{2}_{\rm 1}e^{2}_{\rm 1}\bm{,}
\label{appen2}
\end{split}
\end{equation}

\noindent finally yielding

\begin{equation}
\begin{split}
Q_{xx}-Q_{yy}&=\frac{15}{2}\frac{q}{\left(1+q\right)^{2}} \left(m_{\rm 1}+m_{\rm 2}\right)a^{2}_{\rm 1}e^{2}_{\rm 1} \\
&{\propto}e^{2}_{\rm 1}.
\label{appen6}
\end{split}
\end{equation}

In other words, the strength of the quadrupole in question is proportional to $e^{2}_{\rm 1}$.


\bsp	
\label{lastpage}
\end{document}